\documentclass[aps, pre, showpacs, twocolumn]{revtex4}
\usepackage{MathShorthand}
\usepackage{amsfonts}
\usepackage{amsmath}
\usepackage{empheq}
\usepackage[parfill]{parskip}
\usepackage{graphicx,color}
\usepackage[active]{srcltx}

\begin{document}

\title{Ring states in swarmalator systems}
\date{\today }

\begin{abstract}
Synchronization is a universal phenomenon, occurring in systems as disparate as Japanese tree-frogs and Josephson junctions. Typically, the elements of synchronizing systems adjust the phases of their oscillations, but not their positions in space. The reverse scenario is found in swarming systems, such as schools of fish or flocks of birds; now the elements adjust their positions in space, but without (noticeably) changing their internal states. Systems capable of both swarming and synchronizing,
dubbed swarmalators, have recently been proposed {\cite{o2017oscillators}},
and analyzed in the continuum limit. Here we extend the work in \cite{o2017oscillators} by studying
finite populations of swarmalators, whose phase similarity affects both
their spatial attraction and repulsion. We find ring states,
and compute criteria for their existence and stability. 
Larger populations can form annular distributions, whose density we calculate explicitly. These states may be observable in groups
of Japanese tree frogs, ferromagnetic colloids, and other systems with an interplay
between swarming and synchronization.
\end{abstract}

\author{Kevin P. O'Keeffe}
\affiliation{Senseable City Lab, Massachusetts Institute of Technology, Cambridge, MA 02139, USA,} 

\author{Joep H.M. Evers}
\affiliation{Department of Mathematics and Statistics, Dalhousie University, Halifax, Canada} 

\author{Theodore Kolokolnikov}
\affiliation{Department of Mathematics and Statistics, Dalhousie University, Halifax, Canada} 

\maketitle

\setlength{\parindent}{0.5cm}

%%%%%%%%%%%%%%%%%%%%%%%%%%%%%%%%%%%%%%%%%%%%%%%%%%%%%%%%%%%

\section{Introduction}

Synchronization is a well studied \cite
{pikovsky2003synchronization,strogatz2000kuramoto,winfree,winfree1967biological} phenomenon 
spanning many disciplines. In biology it is seen in discharging pacemaker
cells \cite{peskin1975mathematical, liu1997cellular}, 
coherently flashing fireflies \cite{buck1938synchronous, buck1988synchronous}, and accordantly croaking tree frogs \cite{aihara2007experimental,aihara2008mathematical, Aihara2009}. In chemistry it is seen in the metabolic cycles of yeast cells \cite{aldridge1976cell}, and in
physics, in arrays of Josephson junctions \cite{wiesenfeld1996synchronization}, power grid dynamics \cite{motter2013spontaneous}, 
and even the wobbling of the millenium bridge \cite
{strogatz2005theoretical}.

In synchronizing systems, the dynamic state variables are the oscillators' phases,
whose influence on each other leads to macro-level temporal structures (synchrony). 
A similar effect occurs in swarming  \cite{couzin2002collective,
couzin2003self, couzin2007collective, herbert2016understanding,
sumpter2010collective, mogilner1999non, MEBS, lukeman2010inferring,
bernoff2013nonlocal, FeHuKo11}, a phenomenon as widespread as synchronization, as evidenced by
 flocks of birds, \cite{bialek2012statistical, Ballerini29012008} locust swarms, \cite
{edelstein1998travelling, topaz2008model, buhl2006disorder} bacterial
aggregation, \cite{gdp, msel1,chavy2016local} schools of fish, \cite
{katz2011inferring,barbaro2009discrete}, predator-prey interactions, \cite
{pitcher1983predator, chen2014minimal}, self-assembly \cite
{grzybowski2000dynamic, von2012predicting, von2012soccer,
kolokolnikov2011stability, von2016anisotropic}, and even the vortices of Bose-Einstein
condensates \cite{abo2001observation, neely2010observation, torres2011dynamics,
fine1995relaxation, durkin2000experiments}. Like synchronizing oscillators, the interactions
between swarming particles gives rise to group-level structures. But now
the (dynamic) state variables are the individuals' positions, and the structures formed are spatial.

Viewed this way, swarming and synchronization are strikingly similar. Both are canonical examples of emergent phenomena. Both are dizzyingly pervasive, occurring in far-flung settings like the menstrual 
cycle \cite{mcclintock1971menstrual} and quantum gases \cite{durkin2000experiments}.
Yet in spite of these commonalities, the two fields have developed largely independently.
 In swarming the units are mobile, but do not have internal dynamics. In synchronization 
 the situation is reversed: the oscillators have internal dynamics, but do not move through space.

Recently, however, researchers in both fields have started to study systems with both spatial and internal dynamics. From the swarming side, von Brecht and Uminsky \cite{von2016anisotropic} have endowed
aggregating particles with an internal polarization vector. In the sync community, researchers
have considered mobile oscillators when modeling robotics and biological phenomena \cite{mobile1,stilwell2006sufficient,frasca2008synchronization,fujiwara2011synchronization, buscarino2016interaction}. In these works, however, the coupling between the space dynamics and the phase dynamics is only one way: their phase evolution is influenced by their relative distances, but their relative phases do \textit{not} affect their movements. 

Oscillators whose space dynamics and phase dynamics are bidirectionally coupled have also been considered. The pioneering work was done by Tanaka et al \cite{tanaka2007general,iwasa2011juggling,iwasa2010hierarchical} when 
studying ``chemotactic oscillators", oscillators whose movements and
interactions are mediated by a surrounding chemical. They studied a very general model, from which they derived reduced dynamics using center manifold and phase reductions techniques.
More recent works have been carried out by Starnini et al \cite{starnini2016emergence}, and
O'Keeffe et al \cite{o2017oscillators}, who took a bottom-up approach. They defined minimal, toy models
which enabled greater tractability. The latter called the
elements of their system \textquotedblleft swarmalators" to highlight their
twin identities as swarming oscillators, and to distinguish them from the ``mobile oscillators" of the 
preceding paragraph, whose motion evolves independently of their phase.

Defined this way, swarmalators are, to our knowledge, hypothetical entities. By this we mean there are no real world systems which unequivocally display the required two-way, space-phase coupling. That said, 
there are some promising candidates. For example tree frogs, crickets, and katydids are known to synchronize their calling rhythms with others close to them in space (making the phase dynamics position dependent)  \cite{walker1969acoustic, greenfield1994synchronous}. Perhaps, as some believe \cite{aihara2014spatio}, the relative phases of their calls also affects their movements, which would complete the requisite feedback loop between the space dynamics and the phase dynamics.

Another contender is biological microswimmers, such as bacteria, algae, or sperm. Here the phase variable is associated with the rhythmic wriggling of the swimmer's tail. Since this wriggling both affects, and is affected by, the local hydrodynamic environment, it seems likely that the behavior of neighboring sperm would be coupled. Whether this coupling is truly bidirectional is yet to be determined. That said, there is evidence that sperm, at least, behave this way. As discussed in \cite{yang2008cooperation}, neighboring sperm can synchronize their wriggling, which in turn is thought to enhance their mutual spatial attraction  

Myxobacteria also have the right ingredients to be swarmalators. In this case, the phase variable is an internal, cyclic degree of freedom, which has been theorized to influence their motion, and vice versa \cite{igoshin2001pattern}. The same is true of colloidal Janus particles, where now the phase corresponds to an oscillation about the center of mass (which occurs in response to an external magnetic field). Here again, the physics is such that the oscillations and movements of the particles are mutually dependent on each other, as required of swarmalators \cite{yan2012linking}.

In this work, we contribute to the theoretical study of  swarmalators. We study two realistic modifications of the model defined in \cite{o2017oscillators}. The first is the effect of finite population sizes (in \cite{o2017oscillators} continuum arguments were used), which we show lead to stable ring states. The
second is a change in length scale of the space-phase coupling. In \cite%
{o2017oscillators} this length scale was chosen to be the same as that of the spatial
attraction. However in some swarmalator systems, such as magnetic Janus
particles \cite{yan2012linking} and Japanese tree frogs \cite{Aihara2009},
this space-phase interaction occurs at the length scale of the spatial
repulsion. We here account for this effect by allowing phase similarity to affect
both spatial attraction and spatial repulsion.

%%%%%%%%%%%%%%%%%%%%%%%%%%%%%%%%%%%%%%%%%%%%%%%%%%%%%%%%%%%

\section{The model}

We consider swarmalators confined to move in two spatial dimensions
\begin{align}
\mathbf{ \dot{x}}_{k}   &=  \frac{1}{N}\sum_{j=1}^{N} \Big[ \mathbf{I}_1(\mathbf{x}_{j}-\mathbf{x}_{k})F_1(\theta
_{k}-\theta _{j}) \nonumber \\
& \hspace{2.5 cm} + \mathbf{I}_2(\mathbf{x}_{j}-\mathbf{x}_{k})F_2(\theta _{k}-\theta _{j}) \Big] 
\label{x_eom} \\
 \dot{\theta}_{k} &=\omega _{k}+\frac{K}{N}\sum_{j=1}^{N}H(\theta
_{j}-\theta _{k})G(|\mathbf{x}_{j}-\mathbf{x}_{k}|)  \label{theta_eom}
\end{align}%
for $k=1,\ldots ,N$, where $N$ is the population size and $\mathbf{x}_k  \in \mathbb{R}^2$. $\theta _{k} \in \mathbb{S}^1$ is the phase of the $k$-th swarmalator while its natural frequency is $\omega _{k}$. The spatial attraction and repulsion between swarmalators are represented by $\mathbf{I}_{1}, \mathbf{I}_{2} \in \mathbb{R}^2$. (Depending on the sign of $F_1, F_2$ however, this can change, and $\mathbf{I}_1$ can be repulsive and/or $\mathbf{I}_2$ can be attractive. We discuss when this occurs later). the phase interaction is encoded by $H \in \mathbb{R}$, and the influence of phase similarity on spatial attraction and repulsion is captured by the functions $F_1, F_2 \in \mathbb{R}$. Finally, the function $G \in \mathbb{R}$ represents the influence of spatial proximity on the phase dynamics.  

Consider the following instance of this model: 

%\bes\label{partic}%
\begin{align}
 \mathbf{\dot{x}}_{k}&=\frac{1}{N}\sum_{j\neq i}^{N} \Big(\mathbf{x}_{j}-\mathbf{x}_{k} \Big) \Big( A+J_{1}\cos
(\theta _{j} -\theta _{k})\Big) \nonumber \\
& \hspace{1.5 cm} -\Big( B-J_{2}\cos (\theta _{j}-\theta
_{k})\Big) \frac{\mathbf{x}_{j}-\mathbf{x}_{k}}{|\mathbf{x}_{j}-\mathbf{x}_{k}|^{2}} \label{x_eom_model1} \\
 \dot{\theta}_{k}&=\frac{K}{N}\sum_{j\neq i}^{N}\frac{\sin (\theta
_{j}-\theta _{k})}{|\mathbf{x}_{j}-\mathbf{x}_{k}|^{2}}. \label{theta_eom_model1}
\end{align}
%\ees
\noindent
We choose a linear attraction kernel and power law repulsion, 
as is common in studies of the aggregation model \cite%
{FeHuKo11,EvFeKo2016}, because it simplifies the analysis. 
Specifically, in the absence of space-phase coupling, $J_1 = J_2 = 0$,
this choice of $\mathbf{I}_1, \mathbf{I}_2$  causes swarmalators to form disks of uniform density in space. We note the term $\mathbf{x}_j - \mathbf{x}_k$ indicates the $k$-th swarmalator is attracted to the $j$-th swarmalator only when the term $ (A + J_1 \cos(\theta_j-\theta_k))$ is positive. If the latter term is negative we have the reverse scenario, where the $k$-th swarmalator is \textit{repelled} from the $j$-th swarmalator (similar statements hold for the terms $(\mathbf{x}_{j}-\mathbf{x}_{k}) / |\mathbf{x}_{j}-\mathbf{x}_{k}|^{2}$ and $ B-J_{2}\cos (\theta _{j}-\theta_{k}) ) $. Again for simplicity, we both choose the sine function for $H$, and consider identical swarmalators $\omega_k = \omega$. By a change of reference frame we set $\omega = 0$ without loss of generality. Finally, by rescaling time and space we set $A = B = 1$. Note this implies $(J_1, J_2) \rightarrow ( \tilde{J_1}, \tilde{J_2}) = (J_1 / A B, J_2 / A B ) $, but for notational convenience we drop the tilde notation. This leaves three parameters $(J_{1},J_{2},K)$.

The parameter $K$ measures the strength of the phase coupling. For $K > 0$, the phase
coupling between swarmalators tends to minimize their phase difference,
while for $K < 0$, it tends to maximize it. The parameters $J_1,
J_2 > 0$ measure the extent to which phase similarity influences spatial
attraction and repulsion respectively. For $ 0 < J_1, J_2 < 1$, the functions $F_1$ and
$F_2 $ are strictly positive. Then, the phase similarity enhances just the \textit{magnitude} of $\mathbf{I}_1, \mathbf{I}_2$. However for $J_1, J_2 > 1$, $F_1, F_2$ can change sign (depending on the value of $\theta_j - \theta_k$). As we discussed earlier, this means the functions $\mathbf{I}_1, \mathbf{I}_2$ become repulsive and attractive respectively.

%the sign of these functions can change: 
%if the phases difference between two swarmalators is large enough,
%the attraction between them can become repulsive, and the repulsion between
%them can become attractive.

We remark that $J_2$ does not appear in \cite{o2017oscillators},
which meant phase similarity affected spatial attraction, but not spatial repulsion.
We here include it for greater generality, so that our results may be applied 
to swarmalators whose space-phase coupling occurs on the length scale of the spatial
repulsion, as is the case, for example, for magnetic Janus particles \cite{{janus_rods, yan2012linking}} 
and Japanese tree frogs \cite{aihara2007experimental, Aihara2009}. We also remark that in
\cite{o2017oscillators} $G(\mathbf{|x|}) = 1/|\mathbf{x}|$, but we choose 
$G(\mathbf{|x|}) = 1/|\mathbf{x}|^2$ here because it simplifies the analysis.

%%% Figure 1
\begin{figure*}[t!]
\includegraphics[width= 1.5\columnwidth]{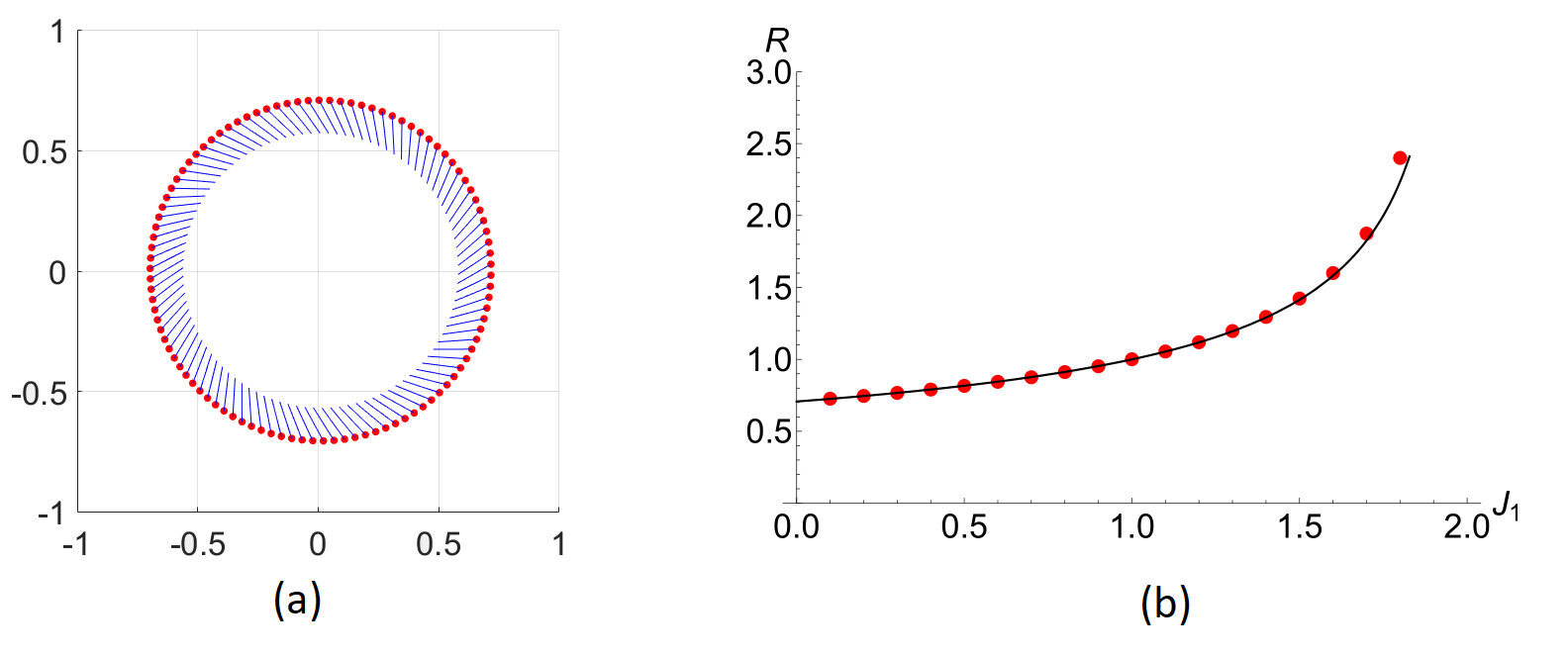}
\caption{(a)\ Scatter plot of a stable ring phase wave state in the $(x,y)$
plane. The phase of each swarmalator is represented by a blue ray, 
and corresponds to the angle the ray makes with the positive $x$-axis. 
As can be seen, in this state the spatial angle $\phi_k = \tan^{-1}(y_k/x_k)$ of each swarmalator
 is correlated with its phase (i.e. $\phi_k = \theta_k + const$). 
Parameter values were $J_{2}=1$, $J_{1}=0,$ $K=-0.003$ and $N=100$.
  (b)\ Radius of ring state versus $J_{1}$. 
  Red dots show simulation results for $J_{2}=1$ and $N=100$. The black curve
shows theoretical prediction \eqref{radius_ring1}. To produce the data for
the plot, we integrated the equations of motion \eqref{x_eom_model1}, \eqref{theta_eom_model1} using
Euler's method until the steady state was
reached.}
\label{fig:ring}
\end{figure*}

%%%%%%%%%%%%%%%%%%%%%%%%%%%%%%%%%%%%%%%%%%%%%%%%%%%%%%%%%%%

\section{Results}

\subsection{Ring phase waves}
Simulations show that for certain parameter values, a stationary state is formed where the swarmalators arrange themselves in a ring centered about the origin, with their phases perfectly correlated with their spatial angle (i.e. $\theta_k = \phi_k + const$, where $\phi_k$ is angle between $\mathbf{x}_k$ and the positive $x$-axis). Accordingly, we call this state the  \textit{ring phase wave} and plot it in Figure~\ref{fig:ring}(a). We now analyze this state. \\

\noindent
\textbf{Existence}. In the ring phase wave state the position and phase of the $k$-th swarmalator are
\begin{align}
\mathbf{x}_{k}& = R \cos{ (2\pi k / N ) } \hat{x} + R \sin{ (2\pi k / N) } \hat{y} \label{ansatz_x} \\
\theta _{k}& = 2 \pi k / N + C \label{ansatz_theta}
\end{align}%
where $R$ is the radius of the ring, $\hat{x}, \hat{y}$ are unit vectors in the $(x,y)$ directions,  $N > 1 $, and the constant $C$ is determined by the initial conditions. After substituting the ansatz %
\eqref{ansatz_x} and \eqref{ansatz_theta} into the equations of motion %
\eqref{x_eom_model1} and \eqref{theta_eom_model1}, and after algebraic manipulation, we derive the following expression for the radius
\begin{equation}
R=\sqrt{\frac{N-1+J_{2}}{N(2-J_{1})}}  \label{radius_ring1}
\end{equation}%
which is valid for any value of the coupling constant $K.$ For large $N$
this becomes $R\sim \sqrt{1/(2-J_{1})}$, independent of $J_{2}$. This
expression for radius of the ring agrees with simulation as shown in Figure~%
\ref{fig:ring}(b). By requiring the argument of the square root be positive, we
see rings which satisfy the ansatz \eqref{ansatz_x}, \eqref{ansatz_theta} exist in the parameter region $\{J_1 < 2, J_2 > 1-N \} \cup \{ J_1 > 2, J_2 < 1-N \}$. \\

%%%%%%%%%%%%%%%%%%%%%%%%%%%%%%%%%%%%%%%%%%%%%%%%%%%%%%%%%%%

\noindent
\textbf{Stability when $ \textbf{K} = \textbf{0} $}. The above analysis proves the existence of ring phase wave, but not their stability, which we here investigate. For simplicity, we start with the case $K  = 0$ so that swarmalators' phases are ``frozen" at the values defined by \eqref{ansatz_theta}. In Appendix B we show that the ring phase wave is stable for $J_{1}\in(J_{1a},2)$ where
\begin{equation}
J_{1a}:=\left\{ 
\begin{array}{c}
2-8\frac{\left( N-1+J_{2}\right) }{\left( N-2\right) ^{2}\left(
1-J_{2}\right) },\ \ \ N\text{ even}, N > 4 \\ 
2-8\frac{\left( N-1+J_{2}\right) }{\left( N-1\right) (N-3)\left(
1-J_{2}\right) },\ \ \ N\text{ odd}, N > 4.%
\end{array}%
\right.  \label{J1a}
\end{equation}

\begin{figure*}[t]
\center
\label{fig:bif}\includegraphics[width=2\columnwidth, height=1.2 \columnwidth]{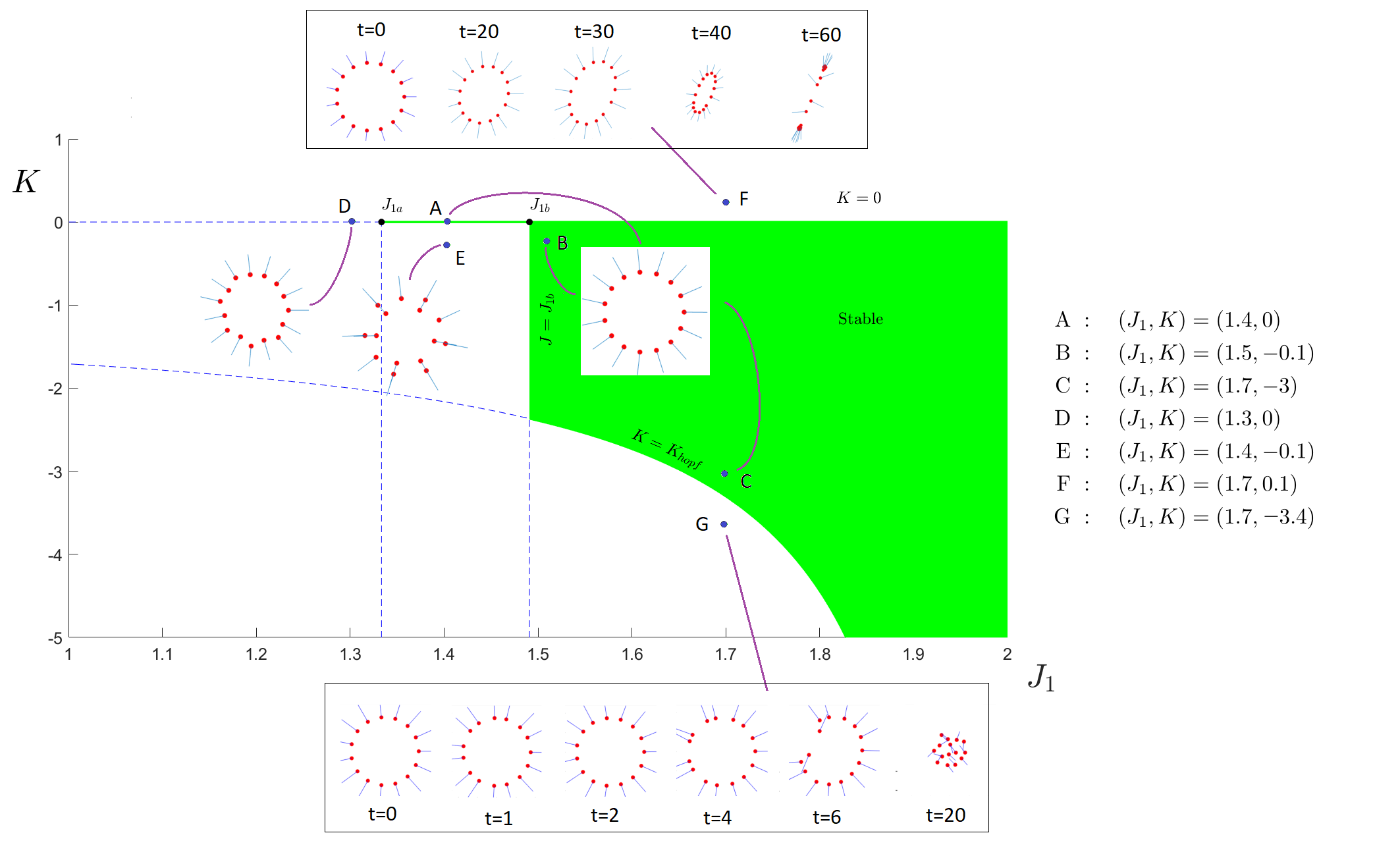}
\caption{Stability diagram for the ring phase wave state in $(J_{1},K)$ space with $%
N=15,J_{2}=0$. Stable regions are indicated with a green color. Inserts show
the solution to Eqs.~\eqref{x_eom_model1} and \eqref{theta_eom_model1} corresponding to parameter values as
shown (A through G) as scatter plots in the $(x,y)$ plane. The phase of each swarmalator
is represented by a blue ray, and corresponds to the angle the ray makes with the positive $x$-axis. 
Initial conditions were taken to be a ring of radius 1,
slightly perturbed. The ring is stable for parameter values A,B,C. }
\label{fig:stab}
\end{figure*}

\noindent
For $J_1 < J_{1a}$ (and $ K = 0$ remember) the ring becomes unstable. However
it does not break up entirely. Instead, it `fattens' slightly, while the phase distribution remaining 
unchanged. This is depicted in snapshot D in Figure \ref{fig:stab}. 
The destabilizing mode in this case is the highest frequency wave number $\left\lfloor N/2\right\rfloor$.

We remark that the case $J_{2}=J_{1}=0$ has a connection to vortex
dynamics. In a classic paper \cite{havelock1931lii}, the stability of
ring configurations of fluid vortices was studied, whose motion is controlled
by the classic Helmholtz equations. It turns out that the
motions of the center of masses of the vortices obey the 
aggregation equation. That is, our governing equations 
 \eqref{x_eom_model1}, \eqref{theta_eom_model1} with $J_1 = J_2 =0$.
 In other words, the vortices swarm. In \cite{chen2013collective} 
 the stability of ring states were studied, and it was found that
  6 or less vortices in the classical vortex equations
are stable, $7$ are neutral (borderline stable/unstable), and 8 or more are
unstable. This is consistent with our result \eqref{J1a}, since $J_{1a} = 0$
at $N=7$ and $J_2 = 0$. \\

%%% Figure 2

%%%%%%%%%%%%%%%%%%%%%%%%%%%%%%%%%%%%%%%%%%%%%%%%%%%%%%%%%%%

\noindent
\textbf{Stability when $\textbf{K} > \textbf{0}$}. When $K$ is positive the swarmalators' phases are 
no longer frozen. Instead, they tend to synchronize with that of their neighbors. This makes ring states unstable. 
A mode-two instability is triggered (which we have determined by numerically computing the eigenvectors), which leads to the \textquotedblleft elliptization\textquotedblright\ of a thin
annulus, as shown in snapshot F of Figure \ref{fig:stab}. This is followed by either
a perfectly synchronous, static crystal formation (equivalent to the ``static sync" state in \cite{o2017oscillators})
or by a blow-up, where the swarmalators escape to infinity. Which of these two states is realized appears (i.e. indicated by numerics) to be parameter dependent (as opposed to depending on initial conditions). Numerics suggest the critical value is at $J_1 \approx 1$ (for $J_2 = 0$) although a theoretical result is lacking. \\

%%%%%%%%%%%%%%%%%%%%%%%%%%%%%%%%%%%%%%%%%%%%%%%%%%%%%%%%%%%

\noindent
\textbf{Stability when $\textbf{K} < \textbf{0}$}. Negative values of $K$ are more interesting.
Now neighboring swarmalators tend to desynchronize their phases. Do rings states persist in this
case? In Appendix B we show they do, provided $J_{1}>J_{1b}$ and $K\in (K_{hopf},0)$ where%
\begin{equation}
J_{1b}=\left\{ 
\begin{array}{c}
2\left( \frac{1}{1-\frac{4}{N^{2}}}\right) -\frac{1}{1-J_{2}}\frac{8}{\left(
N-\frac{4}{N}\right) },\ \ \ N\text{ \ even}, N > 4 \\ 
2\left( \frac{1}{1-\frac{4}{N^{2}-1}}\right) -\frac{1}{1-J_{2}}\frac{8}{
\left( N-\frac{5}{N}\right) },\ \ \ N\text{ odd}, N > 4.
\end{array}
\right. \label{J1b}
\end{equation}
and
\small
\begin{equation}
K_{hopf}=\left\{ 
\begin{array}{c}
-\frac{\left( J_{2}-1\right) \left( -2+J_{1}\right) N^{2}+\left( \left(
-4J_{2}+4\right) J_{1}+8\,J_{2}\right) N+4J_{1}\left( J_{2}-1\right) }{%
N\left( N-4\right) \left( 2-J_{1}\right) }\\ 
-\frac{\left( J_{2}-1\right) \left( -2+J_{1}\right) N^{2}+\left( \left(
-4J_{2}+4\right) J_{1}+8\,J_{2}\right) N+\left( 3J_{2}-3\right)
J_{1}+2J_{2}-2}{\left( N^{2}-4\,N-1\right) \left( 2-J_{1}\right) }
\end{array}%
\right. \label{K_hopf}
\end{equation}
\normalsize

\noindent
where the top equation is for $N$ even, and the bottom is for $N$ odd. As before, these both require $N > 4$.

These instability boundaries are drawn in Figure \ref{fig:stab}. Notice that $J_{1a} < J_{1b}$, so $J_{1b}$
is the critical parameter value when $K < 0 $. Notice also that there are two ways for rings to 
become unstable. The first is by holding $K$ constant, and decreasing $J_1$ below $J_{1b}$ (moving horizontally
in Figure \ref{fig:stab}). This corresponds to a saddle-node bifurcation, and the ring again fattens,
like when $K = 0$. But the similarity (to the scenario when $K=0$) isn't exact; here the phase distribution gets 
distorted (recall it remained unchanged when $K = 0$), as shown in snapshot E of Figure~\ref{fig:stab}.

 %%% Figure 3
\begin{figure*}[t!]
\includegraphics[width=  2\columnwidth]{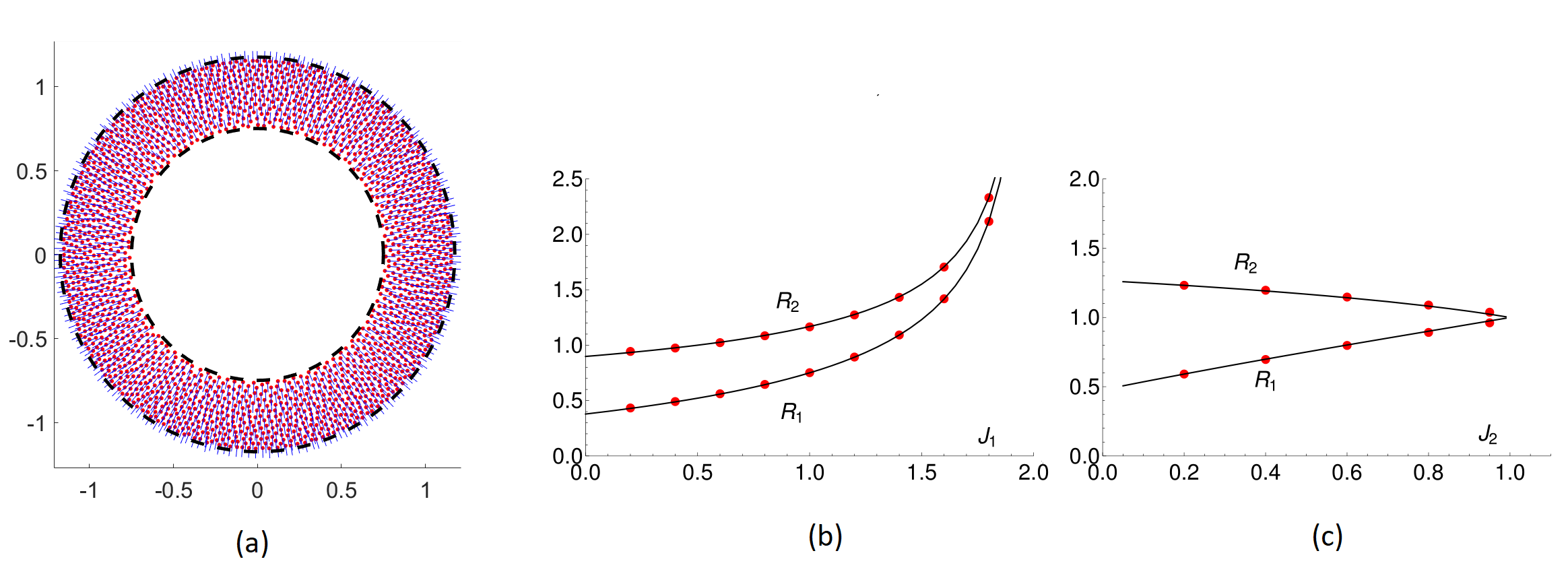}
\caption{The annular phase wave state.
(a): Scatter plot of annular phase wave state in $(x,y)$ plane. The phase of each swarmalator
is represented by a blue ray, and corresponds to the angle the ray makes with the positive $x$-axis. 
Data were collected by solving Eqns~\eqref{x_eom_model1} and \eqref{theta_eom_model1} using the Euler method with $J_1=0.5, J_2=1, K=0$ and $N=2 \times 10^3$ swarmalators.
Asymptotic predictions for the inner and outer radii, as given by the roots of
 \eqref{radii_eqns1} and \eqref{radii_eqns2}, were $R_1 = 0.7504$, $R_2 = 1.16834$,
 and are indicated by dashed curves. Swarmalators were initially placed in a ring
and their initial phases were $\theta_k=\mbox{arg}(x_k)$. 
(b): Comparison of numerics and asymptotic computations of $R_1$ and $R_2$
for $J_2=0.5$ and with varying $J_1$. (c): $J_1=1.0$ and $J_2$ is varied.
}
\label{radii_of_annular_distribution}
\end{figure*}

Rings also become unstable when $J_1$ is held constant, and $K$ is decreased past $K_{hopf} < 0$ (moving vertically
in Figure \ref{fig:stab}). As indicated by the subscript, this leads to a Hopf bifurcation. The ring structure is completely destroyed, 
and a disordered gas-like state forms as
illustrated in snapshot G of Figure~\ref{fig:stab}.  In this state, the swarmalators move
erratically in space and are desynchronized with each other. In the continuum limit 
these movements die out and the ``static async" state reported in
 \cite{o2017oscillators} is achieved, in which the swarmalators form an
 asynchronous disk of uniform density and radius $1$. 
 
We pause to summarize our results so far. We have computed existence and stability criteria for ring states, displayed in the $(J_1, K)$ plane (with $J_2 = 0$ and $N  = 15$) in Figure~\ref{fig:stab}, and discussed the possible bifurcations. We close this section of ring phase wave states by noting some interesting features of the expressions for $J_{1a}, J_{2a}, K_{hopf}$.

The first is their scaling with the population size $N$. For any $N$, it can be shown that $J_{1b}>J_{1a}.$  
Therefore with $K<0$ held fixed, and $J_{1}$ gradually decreased, $J_{2a}$ will be crossed first and the instability changing the phase distribution (snapshot E) will be triggered. When $J_{1a}$ is crossed after this,
the instability shown in snapshot D will be triggered. However as $N\rightarrow \infty ,$ both $J_{1a}\sim
J_{1b}\sim 2-\frac{8}{1-J_{2}}$, which means that the two instabilities happen nearly
simultaneously!

The second interesting feature of the expressions for $J_{1a}, J_{2a}, K_{hopf}$ is that they can
be reversed to find $N(J_1, J_2, K_{hopf})$, allowing us to treat $N$ as a bifurcation parameter. This lets us 
determine  the maximum number of swarmalators in a ring which we define as
\begin{equation}
N_{\max }:=\text{ largest }N\text{ such that }J_{1}>J_{1b}  \label{Nmax}.
\end{equation}%
Then the ring is stable for all $N<N_{\max }$ as long as $K$ is sufficiently
small, namely, $K\in (K_{hopf}(N_{\max }),0].$ When $N$ is large, we can rearrange 
Eq.~\eqref{K_hopf} to obtain
\begin{equation}
N_{\max }\sim \frac{8}{\left( 2-J_{1}\right) \left( 1-J_{2}\right) }.
\label{Nmaxlarge}
\end{equation}%

We restate that the above equation is valid only for large $N$, which means either $0 < 2-J_{1}\ll 1$ or $%
0<1-J_{2}\ll 1$. We see from \eqref{Nmaxlarge} that $N_{\max}$ increases with increasing $J_1$ and $J_2$. Or put another way, swarmalators can form larger rings than regular swarming particles (which have no internal degree of freedom); the inclusion of the phase variable stabilizes the ring state.

The last feature of interest is a special parameter value, $J_2 = 1$, where rings are unusually
stable. To see why, we let $J_2 \rightarrow 1^{-}$ in \eqref{J1a}, \eqref{J1b} and \eqref{K_hopf} and find
\begin{align}
& J_{1a}, J_{2a} \rightarrow - \infty \\
& K_{hopf} \rightarrow \Big\{ 
\begin{array}{c} 
-\frac{8}{(N-4)(2-J_{1})},\ \ \ N\text{ \ even}, N > 4, \; J_2 = 1 \\ 
-\frac{8}{(N-4-1/N)(2-J_{1})},\ \  N\text{ \ odd}, N > 4, \; J_2 = 1%
\end{array}
\end{align}

%\begin{align}
%& J_{1a}, J_{2a} \rightarrow - \infty \\
%& K_{hopf} \rightarrow \left\{ 
%\begin{array}{c} 
%-\frac{8}{(N-4)(2-J_{1})},\ \ \ N\text{ \ even}, N > 4, \; J_2 = 1 \\ 
%-\frac{8}{(N-4-1/N)(2-J_{1})},\ \  N\text{ \ odd}, N > 4, \; J_2 = 1%
%\end{array}
%\end{align}
Consequently, when $J_{2}=1$, $J_1 < 1$ and $K\in
(K_{hopf},0]$ the ring phase wave state is stable for \textit{any} $N$! Furthermore,
its radius is finite, and independent of $N$. This remarkable fact is demonstrated in 
Figure \ref{fig:ring}(a), where a ring of $N=100$ particles is observed to be stable. 

We note that for $J_2 > 1$, simulations show that the particles exhibit
finite-time collisions as $N$ is increased. We therefore restrict our
analysis to the parameter region $J_{2}<1.$  Thus aside from the 
special case $J_{2}=1,$ the ring is stable for $N  < N_{\max}$.
For $N > N_{\max}$ it bifurcates into either the annular phase 
wave state, or the splintered phase wave state, which we discuss next.

%%%%%%%%%%%%%%%%%%%%%%%%%%%%%%%%%%%%%%%%%%%%%%%%%%%%%%%%%%

\subsection{Annular phase waves}

%% Figure 4
\begin{figure*}[t!]
\includegraphics[width= 2 \columnwidth]{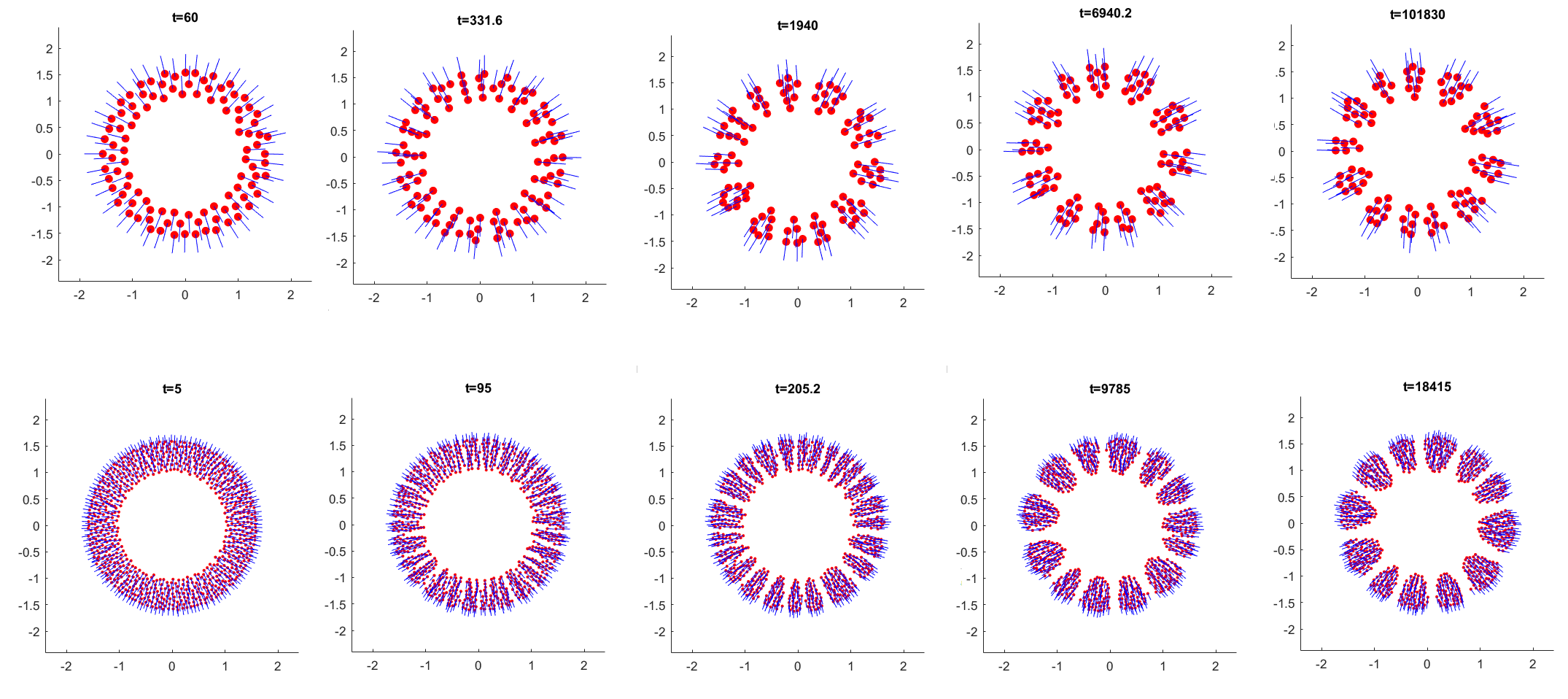}
\caption{Bifurcation of an annulus into a splintered phase wave with 12
clusters. Data were collected by integrating the governing equations \eqref{x_eom_model1},
\eqref{theta_eom_model1} using the Euler method. Parameter values are $J_{1}=1.5,\ J_{2}=0,\ K=-0.05$ and 
$N=100$ (top row) and $N=800$ (bottom row). Swarmalators are illustrated as points in the $(x,y)$ plane, with their phase being represented by a blue ray, and corresponds to the angle the ray makes with the positive $x$-axis. For smaller values of $N$, the system takes longer to equilibrate, and the boundaries between clusters become less well defined.}
\label{splintered}
\end{figure*}

When $N>N_{\max }$ and $K = 0$ the swarmalators form an annular distribution where their spatial angle is perfectly correlated with their phase, plotted in Figure~\ref{radii_of_annular_distribution}(a). This state was reported in \cite{o2017oscillators}, where it was named the ``static phase wave''. To distinguish this state from the ring phase waves of the previous section, we here refer to it as the ``annular phase wave".

We explicitly solve for the density of the annular phase wave in the continuum limit $N \rightarrow \infty$. Let $\rho(\mathbf{x},
\theta, t)$ denote the density of swarmalators, where $\rho(\mathbf{x}, \theta,t) d\mathbf{x} d \theta$ gives the fraction of swarmalators with positions between $\mathbf{x}$ and $\mathbf{x} + d\mathbf{x}$ and phases between $\theta$ and $d \theta$ at time $%
t$. We then use the following ansatz
\begin{equation}
\rho (r,\phi ,\theta ,t)=\frac{1}{2\pi }g(r)\delta (\phi -\theta ),\hspace{%
0.25cm}R_{1}\leq r\leq R_{2} \label{ansatz}
\end{equation}%
where $(r,\phi )$ are polar coordinates and $g(r), R_1, R_2$ are unknown. In Appendix B we solve for $g(r)$ by substituting \eqref{ansatz} into the continuity equation and deriving an integral equation for $g(r)$. We then reduce this integral equation to  second order ODE, whose solution is
\begin{equation}
g(r) = C_1 r^{-\frac{1}{\sqrt{1-J_2}}-2}+C_2 r^{\frac{1}{\sqrt{1-J_2}}-2}+\frac{6}{3-4J_2}
\end{equation}%
where $C_{1},C_{2}$ are complicated expressions involving $%
R_{1},R_{2},J_{1},J_{2}$ given by Eq.~\eqref{C1} and Eq.~\eqref{C2}. Note this is valid for $J_2 \neq 3/4$. At this parameter value, $g(r)$ takes a different functional form, which we display and discuss in Appendix B.

We also derive implicit equations for the inner and outer radii $R_1, R_2$ in terms of $J_1, J_2$
\begin{align}
&h_1(R_1, R_2, J_1, J_2) = 0  \label{radii_eqns1} \\
& h_2(R_1, R_2, J_1, J_2) = 0  \label{radii_eqns2}.
\end{align}
\noindent
where $h_1, h_2$ are complicated expressions given by Eq.~\eqref{h1} and Eq.~\eqref{h2}. We solved these using Mathematica. The results are shown in Figure~\ref{radii_of_annular_distribution}(b) and Figure~\ref{radii_of_annular_distribution}(c), which agree well with numerics

Notice in Figure~\ref{radii_of_annular_distribution} that $R_1 \rightarrow R_2$ as $J_1 \rightarrow 2$ in panel (b) and $J_2 \rightarrow 1$ in panel (c), indicating the morphing of the annular phase wave into the ring phase wave state. We analytically confirm $J_{2c} = 2$ by substituting $R_1 = R_2$ into \eqref{radii_eqns1}. The result is
\begin{equation}
(3 - 4 J_2) (-1+ J_2+\sqrt{1-J_2}) R_2^{\frac{2}{\sqrt{1-J_2}}} = 0 \label{tempB}
\end{equation}
\noindent 
From this we see $-1+ J_2+\sqrt{1-J_2} = 0$ which gives
\begin{equation}
J_{2c} = 1.
\end{equation}
\noindent 
Note \eqref{tempB} is only valid for $J_2 \neq 3/4$, a property inherited from the expression for $g(r)$ (see Appendix B). We confirm the $J_{1c}$ value similarly; we substituted $R_1 =
R_2 - \delta$ into \eqref{radii_eqns2} and took a
series expansion for small $\delta$ leading to
\begin{align}
\left(J_1-2\right) \left(4 J_2+3\right) \left(-J_2+\sqrt{J_2+1}-1\right) \nonumber \\
\times \left(\frac{J_2+\sqrt{\left(J_2+1\right){}^2}+1}{\delta }\right){}^{\frac{2}{%
\sqrt{J_2+1}}} = 0
\end{align}

\noindent from which we see
\begin{equation}
J_{1c} = 2.
\end{equation}

We close by distilling our results. We explicitly solved for the density in the annular phase wave state, and showed it exists in the parameter region $0 < J_{1} < 2,  0 < J_{2}<1$. As the extremal edges of this region are approached, the annulus gets thinner and 
thinner until the ring phase wave is achieved right at the boundary $J_1 = 2$ or $J_2 = 1$. When $J_1 = 2$, the radius of the ring approaches $\infty$, whereas when $J_{1}\rightarrow 2^{-}$ it remains finite. Note that we have only proved the \textit{existence} of the annular
phase wave here, and make no claims about its stability. Numerics indicate that it is stable, but a proof is beyond the scope of the present work.

%%%%%%%%%%%%%%%%%%%%%%%%%%%%%%%%%%%%%%%%%%%%%%%%%%%%%%%%%%%

\subsection{Splintered phase wave.}
In the above section we showed that when $K = 0$ and $N > N_{\max}$, the ring phase wave bifurcates into 
the annular phase wave. For $ K < 0$, they bifurcate into a new state called the \textit{splintered phase wave}, previously reported in \cite{o2017oscillators}. Here, the ring `splinters' into disconnected clusters of distinct phase. Within each cluster, swarmalators `quiver', executing small cycles in both position and phase about their mean values. We showcase the evolution of this state from the annular phase wave in Figure~\ref{splintered}. 

This non-stationary behavior makes analysis difficult, and we were unable to 
construct the state or determine its stability. We were however able to heuristically
 find an upper bound for the number of clusters that form. We did this
by leveraging our analysis for the ring states: we naively pictured
each cluster as a single particle, which lets us reimagine the splintered phase 
wave state as a ring state. We then use our previous analysis to estimate
$N_{\max}$ given by \eqref{Nmax}. For example, for parameter values used in Figure~
\ref{splintered}, $N_{\max }=15,$ whereas the number of observed clusters is 12 or 13. Simulations
at other parameter values have the same behavior.

%%%%%%%%%%%%%%%%%%%%%%%%%%%%%%%%%%%%%%%%%%%%%%%%%%%%%%%%%%%

\subsection{Genericity.}
So far our analysis has been for the instance \eqref{x_eom_model1}, \eqref{theta_eom_model1} of the model \eqref{x_eom}, \eqref{theta_eom}. We here check if the phenomena we found are generic to the model, rather than specific to the instance of the model. We do this by exploring the effects of different functional forms for $\mathbf{I}_1, \mathbf{I}_2, F, G$. We study three such choices, listed below. In all cases we found the same states enumerated in Figure~\ref{fig:stab}. We exhaustively show these states for all three choices of interaction function in Figure~\ref{genericity} in Appendix C.

\begin{align}
\mathbf{I}_1, \mathbf{I_2}, G, H &= \frac{\mathbf{x}}{|\mathbf{x}|^2}, \frac{\mathbf{x}}{|\mathbf{x}|^4}, \frac{1}{|\mathbf{x}|}, \sin \theta \label{gener1} \\
\mathbf{I}_1,  \mathbf{I}_2, G, H &= \mathbf{x} e^{- |\mathbf{x}|}, \frac{\mathbf{x}}{|\mathbf{x}|^2}, \frac{1}{|\mathbf{x}|}, \sin \theta \label{gener2} \\
\mathbf{I}_1,  \mathbf{I}_2, G, H &= \mathbf{x}, \frac{\mathbf{x}}{|\mathbf{x}|^2}, \frac{e^{-|\mathbf{x}|}}{|\mathbf{x}|^2}, \sin \theta \label{gener3}
\end{align}

\begin{figure}[t!]
\includegraphics[width= 0.7 \columnwidth]{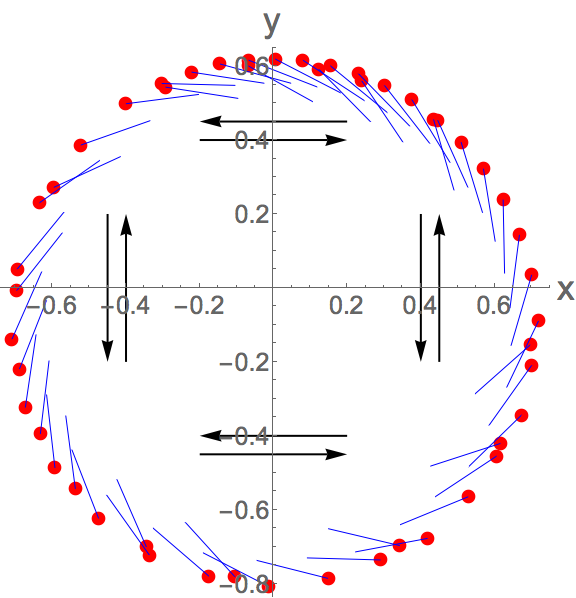}
\caption{Ring state in the presence of heterogenous natural frequencies $\omega_k = \omega_0 + (2 k \omega_0)/(N-1)$ with $\omega_0 = 0.2$, $N = 50$ and $k = 1,2,3 \dots$. We used the Euler method with stepsize $dt = 0.1$. The ring distribution remains, but swarmalators are no longer stationary; they split into counter rotating (in both space and phase) groups, as indicated by the black arrows. This shear-like flow was reported in \cite{o2017oscillators}, where it was named the active phase wave state. }
\label{omega}
\end{figure}

We were also curious if the ring state would persist in the presence of heterogeneity. To this end, we imbued swarmalators with  natural frequencies $\omega_k$ linearly spaced on $[-\omega_0, \omega_0]$ (recall so far we have considered identical swarmalators $\omega_i = \omega = 0$ -- the zero value achieved by a change of reference). Simulations show the ring distribution persists, but now the swarmalators split into counter rotating groups (which follows from the fact that $\langle \dot{\theta}_i \rangle  = \langle \dot{x}_i \rangle = 0$ in our model). That is, individual swarmalators execute circular motion in both space and phase, with the overall \textit{density} of swarmalators remaining constant. This state is equivalent to the active phase wave reported in \cite{o2017oscillators}, with the inner and outer radii of the annular being the same. Figure~\ref{omega} displays the state in the $(x,y)$ plane. A theoretical understanding of this state is lacking (aside from the trivial result that the radius of the ring is still given by \eqref{radius_ring1}), and is left for future work.

%% Figure 5

%%%%%%%%%%%%%%%%%%%%%%%%%%%%%%%%%%%%%%%%%%%%%%%%%%%%%%%%%%

\section{Discussion}
We studied the stability of ring states in swarmalator systems with both phase dependent attraction and phase dependent repulsion. We analytically computed criteria for their existence and stability, which were valid for all population sizes $N$. We found that in general (even for $K$ sufficiently small and negative) ring states are stable 
for sufficiently small populations $ N < N_{\max}$. For $N>N_{\max}$, they bifurcate into either the
annular phase wave or splintered phase wave state. We constructed the former state  in
the continuum limit $N \rightarrow \infty$, but its stability remains an open problem. We were unable
to construct the latter state, or determine its stability, and so these are also open problems. We were 
however able to heuristically derive an upper bound for the number of synchronous clusters which comprise the state.

Ring states have been previously studied in `regular' swarming systems, where
particles have a position $x_k$ but no internal phase $\theta_k$. They were first 
shown to be stable in two dimensions \cite{kolokolnikov2011stability, bertozzi2015ring},
and later in three \cite{von2012predicting, von2012soccer}. The general case of $n$
dimensions was completed in \cite{balague2013dimensionality}, where the authors
showed that the formation of rings depends on the strength of the near-field repulsion
(more precisely, they show the support of the local minimizer of the interaction potential
 has Hausdorff dimension greater than or equal to the strength of the repulsion at the origin). 
This means rings can only form when the repulsion between two particles is bounded (i.e. no hard
shell repulsion). Interestingly, we have demonstrated this is not true for swarmalators: our  
repulsion term was hard shelled (see Eq.~\eqref{x_eom_model1}), yet we proved rings 
are stable for certain parameter values (detailed in Figure~\ref{fig:stab}). 

A similar result is found in anisotropic swarming systems, where the particles now have an additional state
variable such as an orientation or a heading vector. For example, von Brecht and Uminsky \cite{von2016anisotropic} used an anisotropic version of the aggregation equation in 3D to explore the effects of polarization on molecular structures, and found that anisotropy enhanced the stability of `blackberries' - shell like structures found in biochemical contexts. This echoes our finding that the inclusion of a phase in swarming systems stabilizes ring states.
It seems the addition of a circular state variable (for swarmalators an internal phase, and for swarming particles an orientation/heading) stabilizes structures of low co-dimension (rings/shells). Rigorously justifying this claim is an interesting open problem; perhaps an extension of the techniques used in \cite{balague2013dimensionality} could prove fruitful.

An apposite future goal would be to find or manufacture real-world realizations of the states here studied. States similar to the rings and static phase wave have been realized in ferromagnetic colloids confined to liquid-liquid interfaces. So called `asters' consist of annular structures of particles whose magnetic dipole vectors correlate with their spatial angle \cite{snezhko2011magnetic}, as happens in the ring and static phase wave states studied here. Ring-like states are found in groups of Japanese tree frogs, who congregate along edges of paddy fields \cite{aihara2008mathematical}. The phase distribution is however different to that found here; instead neighboring frogs are perfectly out of phase with each other. Full phase waves are yet to be discovered.

There are also theoretical avenues for future work within our proposed model of swarmalators. For instance we considered motion in just two spatial dimensions. While there are some physical systems where this type of motion is realized, such as certain active colloids \cite{pohl2014dynamic} or sperm -- which are often attracted to the surface of liquids \cite{maude1963non} -- this was mostly for mathematical convenience. The more realistic case of motion in three spatial dimensions would be interesting to explore. For instance, 3D analogues of the states found in 2D were reported in \cite{o2017oscillators}, but their stability wasn't analyzed. Moreover, finite populations sizes were unexplored. Perhaps the analysis in \cite{von2016anisotropic} would be helpful in answering these questions.

Other extensions include adding heterogeneity in the coupling parameters $K, J_1, J_2$, and the natural frequencies $\omega_k$, or considering delayed or noisy interactions. Less trivial phase dynamics could also be interesting. As we stated, the choice of $H(\theta) = \sin(\theta)$ was inspired by the Kuramoto model \cite{kuramoto}, but leads to trivial phenomena in the $ K > 0$ plane (total synchrony). Perhaps using the more realistic Winfree model \cite{winfree}, which has richer phase dynamics, would lead to more interesting swarmalator phenomena when $K$ is positive. 

%%%%%%%%%%%%%%%%%%%%%%%%%%%%%%%%%%%%%%%%%%%%%%%%%%%%%%%%%%

\section{Acknowledgments}

Research supported by NSF Grant Nos DMS-1513179 and CCF-1522054 (K.P.O.) and NSERC Discovery Grant No. RGPIN-33798 and Accelerator Supplement Grant No. RGPAS/461907 (T.K. \& J.H.M.E).

%%%%%%%%%%%%%%%%%%%%%%%%%%%%%%%%%%%%%%%%%%%%%%%%%%%%%%%%

\appendix

\section{Stability of ring phase wave}
Here, we develop the stability theory for ring states of the swarmalator
model defined in the main text, using techniques similar to those developed
in \cite{kolokolnikov2011stability, kolokolnikov2014tale, bertozzi2015ring,
albi2014stability}. It is convenient to use complex notation to describe the
ring phase wave state. We thus identify the real, two dimensional, vector $\mathbf{x}_{k} = (x_k^{(1)}, x_k^{(2)})$ as a point in the complex plane (so that $x_k^{(1)}$ is real part of the complex number, and $x_k^{(2)}$ is the imaginary part). To remind ourselves that $\mathbf{x}_k$ is now a complex number, we drop the bold notation hereafter.

We first consider a more general model of
the form{\ 
\begin{align}
x_{k}^{\prime }& =\sum_{j}f\left( \left\vert x_{k}-x_{j}\right\vert
^{2}\right) \left( x_{k}-x_{j}\right)  \notag \\
&+\sum_{j}\cos \left( \theta _{k}-\theta _{j}\right) h\left( \left\vert
x_{k}-x_{j}\right\vert ^{2}\right) \left( x_{k}-x_{j}\right)  \label{gen1} \\
\theta _{k}^{\prime }& =\sum_{j}\sin \left( \theta _{k}-\theta _{j}\right)
g\left( \left\vert x_{k}-x_{j}\right\vert ^{2}\right) .  \label{gen2}
\end{align}%
} The model defined by Eqns \eqref{x_eom_model1} \eqref{theta_eom_model1}
then corresponds to the specific choice{%
\begin{equation}
f(r)=\frac{1}{r}-1;\ \ \ h(r)=-\frac{J_{2}}{r}-J_{1},\ \ \ \ g(r)=-\frac{K}{r%
}.  \label{fgh}
\end{equation}%
} The ring phase wave steady state is given by\ {\ 
\begin{align*}
x_{k}& =Rz^{k},\ \ \ \text{where}\ \ \ z:=\exp \left( 2\pi i/N\right) , \\
\theta _{k}& =2\pi k/N
\end{align*}%
}where $R$ is the ring radius. This ansatz satisfies Eq.~(\ref{gen2}) for
any $R$ whereas (\ref{gen1}) is satisfied if and only if 
\begin{align}
&\sum_{l\neq 0}f\left( R^{2}\left\vert 1-z^{l}\right\vert ^{2}\right) \left(
1-z^{l}\right)  \notag \\
&+\sum_{l\neq 0}h\left( R^2 \left\vert 1 -z^{l}\right\vert ^{2}\right) \cos
\left( 2\pi l/N\right) \left( 1-z^{l}\right) =0.  \label{10:02}
\end{align}%
which gives an expression for $R.$ For the specific choice (\ref{fgh}),
using the identities%
\begin{equation}
\sum_{l\neq 0}\frac{1}{1-z^{l}}=\frac{N-1}{2},\ \sum_{l\neq 0}\frac{%
z^{l}+z^{-l}}{1-z^{-l}}=-1,  \label{10:03}
\end{equation}%
Eq.~(\ref{10:02})\ reduces to Eq.~\eqref{radius_ring1}.

We now consider the perturbations,%
\begin{equation*}
x_{k}(t)=Rz^{k}+u_{k}(t);\ \ \ \theta _{k}=2\pi k/N+v_{k}(t).
\end{equation*}

Substituting into the governing equations and linearizing gives\ 
\begin{widetext}
\begin{align*}
u_{k}^{\prime }& =\sum_{j}\left[ f^{\prime }\left( \left\vert
x_{k}-x_{j}\right\vert ^{2}\right) +\cos \left( \theta _{k}-\theta
_{j}\right) h^{\prime }\left( \left\vert x_{k}-x_{j}\right\vert ^{2}\right) %
\right] \left( x_{k}-x_{j}\right) ^{2}\left( \overline{u_{k}-u_{j}}\right)
-J\sin \left( \theta _{k}-\theta _{j}\right) h\left( \left\vert
x_{k}-x_{j}\right\vert ^{2}\right) \left( x_{k}-x_{j}\right) \left(
v_{k}-v_{j}\right) \\
& +\sum_{j}\left[ 
\begin{array}{c}
f\left( \left\vert x_{k}-x_{j}\right\vert ^{2}\right) +f^{\prime }\left(
\left\vert x_{k}-x_{j}\right\vert ^{2}\right) \left\vert
x_{k}-x_{j}\right\vert ^{2}+\cos \left( \theta _{k}-\theta _{j}\right)
h\left( \left\vert x_{k}-x_{j}\right\vert ^{2}\right) \\ 
+\cos \left( \theta _{k}-\theta _{j}\right) h^{\prime }\left( \left\vert
x_{k}-x_{j}\right\vert ^{2}\right) \left\vert x_{k}-x_{j}\right\vert ^{2}%
\end{array}%
\right] \left( u_{k}-u_{j}\right) \\
&
\end{align*}%
and {\ 
\begin{equation*}
v_{k}^{\prime }=\sum_{j}\sin \left( \theta _{k}-\theta _{j}\right) g^{\prime
}\left( \left\vert x_{k}-x_{j}\right\vert ^{2}\right) \left\{ \left(
x_{k}-x_{j}\right) \left( \overline{u_{k}-u_{j}}\right) +\left( \overline{%
x_{k}-x_{j}}\right) \left( u_{k}-u_{j}\right) \right\} +\sum_{j}\cos \left(
\theta _{k}-\theta _{j}\right) g\left( \left\vert x_{k}-x_{j}\right\vert
^{2}\right) \left\{ v_{k}-v_{j}\right\} .
\end{equation*}%
Following \cite{kolokolnikov2011stability, kolokolnikov2014tale,
bertozzi2015ring}, we use the self-consistent ansatz\ 
\begin{align*}
u_{k}(t)& =A(t)z^{mk+k}+\bar{B}(t)z^{-mk+k} \\
v_{k}& =C(t)z^{mk}+\bar{C}(t)z^{-mk}.
\end{align*}%
After much algebra, and collecting like-terms in }$z^{mk}$ and $z^{-mk}$, we
obtain a 3x3 linear system for each mode $m${\ 
\begin{equation}
\left( 
\begin{array}{c}
A^{\prime } \\ 
B^{\prime } \\ 
C^{\prime }%
\end{array}%
\right) =\left( 
\begin{array}{ccc}
M_{11} & M_{12} & M_{13} \\ 
M_{21} & M_{22} & M_{23} \\ 
M_{31} & M_{32} & M_{33}%
\end{array}%
\right) \left( 
\begin{array}{c}
A \\ 
B \\ 
C%
\end{array}%
\right)
\end{equation}%
where%
\begin{align*}
M_{11}& =\sum \left[ 
\begin{array}{c}
f\left( R^{2}\left\vert 1-z^{l}\right\vert ^{2}\right) +f^{\prime }\left(
R^{2}\left\vert 1-z^{l}\right\vert ^{2}\right) R^{2}\left\vert
1-z^{l}\right\vert ^{2} \\ 
+\cos \left( \frac{2\pi l}{N}\right) \left( h\left( R^{2}\left\vert
1-z^{l}\right\vert ^{2}\right) +h^{\prime }\left( R^{2}\left\vert
1-z^{l}\right\vert ^{2}\right) R^{2}\left\vert 1-z^{l}\right\vert ^{2}\right)%
\end{array}%
\right] \left( 1-z^{(m+1)l}\right) \\
M_{12}& =\sum \left[ f^{\prime }\left( R^{2}\left\vert 1-z^{l}\right\vert
^{2}\right) +\cos \left( \frac{2\pi l}{N}\right) h^{\prime }\left(
R^{2}\left\vert 1-z^{l}\right\vert ^{2}\right) \right] R^{2}\left(
1-z^{l}\right) ^{2}\left( 1-z^{(m-1)l}\right) \\
M_{13}& =\sum h\left( R^{2}\left\vert 1-z^{l}\right\vert ^{2}\right) \sin
\left( 2\pi l/N\right) R\left( 1-z^{l}\right) \left( 1-z^{ml}\right)
\end{align*}%
} {\ and%
\begin{align*}
M_{21}& =M_{12} \\
M_{22}& =\sum \left[ 
\begin{array}{c}
f\left( R^{2}\left\vert 1-z^{l}\right\vert ^{2}\right) +f^{\prime }\left(
R^{2}\left\vert 1-z^{l}\right\vert ^{2}\right) R^{2}\left\vert
1-z^{l}\right\vert ^{2} \\ 
+\cos \left( \frac{2\pi l}{N}\right) \left( h\left( R^{2}\left\vert
1-z^{l}\right\vert ^{2}\right) +h^{\prime }\left( R^{2}\left\vert
1-z^{l}\right\vert ^{2}\right) R^{2}\left\vert 1-z^{l}\right\vert ^{2}\right)%
\end{array}%
\right] \left( 1-z^{(m-1)l}\right) \\
M_{23}& =\sum \sin \left( \frac{2\pi l}{N}\right) h\left( R^{2}\left\vert
1-z^{l}\right\vert ^{2}\right) R\left( 1-z^{-l}\right) \left( 1-z^{ml}\right)
\end{align*}%
and%
\begin{align*}
M_{31}& =\sum -\sin \left( 2\pi l/N\right) g^{\prime }\left( R^{2}\left\vert
1-z^{l}\right\vert ^{2}\right) \left\{ R\left( 1-z^{-l}\right) \left(
1-z^{(m+1)l}\right) \right\} \\
M_{32}& =\sum -\sin \left( 2\pi l/N\right) g^{\prime }\left( R^{2}\left\vert
1-z^{l}\right\vert ^{2}\right) \left\{ R\left( 1-z^{l}\right) \left(
1-z^{(m-1)l}\right) \right\} \\
M_{33}& =\sum \cos \left( 2\pi l/N\right) g\left( R^{2}\left\vert
1-z^{l}\right\vert ^{2}\right) \left( 1-z^{ml}\right) .
\end{align*}%
where all sums are over }$l=1\ldots N-1$. Specializing to (\ref{fgh}), we
use the following key identity:{%
\begin{equation*}
\sum_{l=1}^{N-1}\frac{z^{ml}}{\left( 1-z^{l}\right) ^{2}}=\left\{ 
\begin{array}{c}
\frac{1}{12}+\frac{1}{24}N^{2}-\frac{1}{2}\left( m-1-N/2\right) ^{2},\ \ \ \
\ m\in \left( 1,N-1\right) \\ 
-\frac{1}{12}\left( N-5\right) (N-1),\ \ \ \ \ m\equiv 0%
\end{array}%
\right.
\end{equation*}%
}

The expressions for $M$ then become,
\begin{equation*}
M=\left[ 
\begin{array}{ccc}
-N+J_{1}\frac{N}{2} & \frac{\left( N-3\right) (1-J_{2})}{2R^{2}} & 0 \\ 
\frac{\left( N-3\right) (1-J_{2})}{2R^{2}} & \frac{N}{2}J_{1}-N & 0 \\ 
0 & 0 & 0%
\end{array}%
\right] ,\ \ \ \ m=0
\end{equation*}%
\begin{equation*}
M=\left[ 
\begin{array}{ccc}
-N & \frac{\left( N-4\right) (1-J_{2})}{R^{2}} & i\frac{N}{2}\left( 2RJ_{1}+%
\frac{J_{2}}{R}\right)  \\ 
\frac{\left( N-4\right) (1-J_{2})}{R^{2}} & 0 & 0 \\ 
-i\left( N-2\right) \frac{K}{2R^{3}} & 0 & \frac{K}{2R^{2}}%
\end{array}%
\right] ,\ \ \ \ m=1
\end{equation*}%
\begin{equation*}
M=\left[ 
\begin{array}{ccc}
-N & \frac{3\left( N-5\right) }{2}\frac{(1-J_{2})}{R^{2}} & i\frac{N}{2}%
\left( RJ_{1}+\frac{J_{2}}{R}\right)  \\ 
\frac{3\left( N-5\right) }{2}\frac{(1-J_{2})}{R^{2}} & \frac{N}{2}J_{1}-N & 
-i\frac{N}{2}\left( 2RJ_{1}+\frac{J_{2}}{R}\right)  \\ 
-iK\frac{\left( N-3\right) }{R^{3}} & iK\frac{\left( N-2\right) }{2R^{3}} & -%
\frac{K}{2R^{2}}\left( N-4\right) 
\end{array}%
\right] ,\ \ \ \ m=2
\end{equation*}%
\bigskip 

For $m\in \left( 2,N-2\right) ,$ we have, 
\begin{equation*}
M=\left[ 
\begin{array}{ccc}
-N & \frac{\left( m-1\right) \left( -m+N-1\right) \left( -J_{2}+1\right) }{%
2R^{2}} & i\frac{N}{2}\left( RJ_{1}+\frac{J_{2}}{R}\right) \\ 
\frac{\left( m-1\right) \left( -m+N-1\right) \left( -J_{2}+1\right) }{2R^{2}}
& -N & -i\frac{N}{2}\left( RJ_{1}+\frac{J_{2}}{R}\right) \\ 
-K\frac{i}{2R^{3}}\left( N-m-1\right) m & K\frac{i}{2R^{3}}\left( m-1\right)
\left( N-m\right) & -\frac{K}{2R^{2}}\left( N(m-1)-m^{2}\right)%
\end{array}%
\right] .
\end{equation*}%
It turns out that the modes $m=0,1,2$ are stable in the relevant regimes so
we do not examine them here. We have checked this analytically for $K  = 0$, but for $K \neq 0$, we were only able to do this numerically. As we will show however, the expression for $m  \in (2,N-2)$, leads to closed form expressions for critical parameters -- values that match simulations -- so we confine our attention there hereafter. The above matrix (i.e. for $ m \in (2,N-2)$) has the following form.
\begin{equation}
M=\left[ 
\begin{array}{ccc}
a & b & ic \\ 
b & a & -ic \\ 
iKd & iKe & Kf%
\end{array}%
\right]
\end{equation}%
where%
\begin{eqnarray}
a &=&-N,\ \ b=\frac{\left( m-1\right) \left( -m+N-1\right) \left(
-J_{2}+1\right) }{2R^{2}},\ \ c=\frac{N}{2}\left( RJ_{1}+\frac{J_{2}}{R}%
\right) \\
d &=&\frac{-\left( N-m-1\right) m}{2R^{3}},\ \ \ e=\frac{\left( m-1\right)
\left( N-m\right) }{2R^{3}},\ \ \ f=\frac{m^{2}-N(m-1)}{2R^{2}}.  \notag
\end{eqnarray}%
Computing the characteristic polynomial, we find that one of the eigenvalues
is given by%
\begin{equation}
\lambda _{0}=a+b  \label{10:04}
\end{equation}%
while the other two are roots of the quadratic 
\begin{equation}
K\left( f(a-b)+c(d-e)\right) +\lambda \left( b-a-Kf\right) +\lambda ^{2}=0.
\label{10:05}
\end{equation}
\end{widetext}

We remind the reader that these expressions are for $m \in (2,N-2)$. This
requires $N > 4$. Thus, the following analysis holds only when this
condition is met.

From the expressions of the eigenvalues we deduce the instabilities that can
occur. There are three types:\ either (\ref{10:04})\ crosses through zero, (%
\ref{10:05})\ crosses through zero, or (\ref{10:05})\ exhibits a Hopf
bifurcation. These three possibilities correspond to $a+b=0,\ \ K\left(
f(a-b)+c(d-e)\right) =0$, and $b-a-Kf=0\,\ $(with $K\left(
f(a-b)+c(d-e)\right) <0$)$,\ $respectively.

Further analysis shows that the ring is unstable with respect to mode $m=2$
whenever $K>0,$ regardless of the values of $J_{1},J_{2}.$ Hence we ignore
this boring part of parameter space and consider only the region $K\leq 0.$
It turns out that the most unstable mode corresponds to the highest mode $%
m=\left\lfloor N/2\right\rfloor $. With this choice of $m,$ let $J_{1a}$ be
the value of $J_{1}$ such that $a+b=0$, and let $J_{1b}$ be the value $J_{1}$
such that $f(a-b)+c(d-e)=0.$ Finally, let $K_{hopf}$ be the value of $K$ for
which $b-a-Kf=0.$ These values are given by \eqref{J1a}, \eqref{J1b}, and %
\eqref{K_hopf} in the main text respectively. Further analysis shows that $%
J_{1a}<J_{1b}.$ (Note, the swarmalators execute oscillations in both space and phase after the hopf bifurcation) 

The stability diagram is illustrated in Figure \ref{fig:stab}. Suppose that $%
K\leq 0.$ Then for $J_{1}$ below $J_{1a},$ the ring is unstable with respect
to spatial perturbation. For $J_{1a}<J_{1}<J_{1b},$ the ring is unstable
with respect to a mixture of spatial and phase perturbations, when $K<0,$
but is stable when $K=0.$ Finally, the ring is fully stable if $J_{1b}<J_{1}$
as long as $K_{hopf}<K<0.$ This stability region is indicated in green in
Figure \ref{fig:stab}.

%%%%%%%%%%%%%%%%%%%%%%%%%%%%%%%%%%%%%%%%%%%%%%%%%%%%%%%%

\section{Density of annular phase wave state}
The density of swarmalators in the annular phase wave state (best expressed in polar coordinates) in given by
\begin{align}
\rho(r, \phi, \theta) &= \frac{1}{2 \pi} g(r) \delta(\phi - \theta), \hspace{%
0.25 cm} R_1 \leq r \leq R_2  \label{density_ansatz} \\
&= 0, \hspace{0.25 cm} \text{elsewhere}
\end{align}
\noindent where $r_k, \phi_k$ is the radial position and
spatial angle of the $k$-th swarmalator, and $g(r),R_1, R_2$ are unknowns to be solved for. We first solve for $g(r)$, which in turn lets us solve for $R_1, R_2$.

%%%%%%%%%%%%%%%%%%%%%%%%%%%%%%%%%%%%%%%%%%%%%%%%%%%%%%%%

\subsection{Find radial density $g(r)$}
Swarmalators are stationary (in both space and phase) in the annular phase wave state:
\begin{equation}
\underline{v} \equiv \underline{0} \label{vel}
\end{equation}
where we have introduced the ``underline" notation $\underline{v} = (\mathbf{v_x}, v_{\theta})$ (so that $\underline{v} \in \mathbb{R}^3$, $\mathbf{v_{x}} \in \mathbb{R}^2$ and $v_{\theta} \in \mathbb{R}$). By applying the divergence operator to \eqref{vel} we generate another equation
\begin{equation}
\nabla . \underline{v} \equiv 0. \label{div_vel}
\end{equation}
Equations \eqref{vel} and \eqref{div_vel} let us solve for $g(r)$, as we will now show. \\

%\begin{align}
%\mathbf{v_x} &= \int \Bigg[ (\mathbf{x^{\prime}} - \mathbf{x}) \Big( 1 + J_1 \cos(\theta^{\prime} - \theta)\Big) \nonumber \\
% &- \Big( 1 - J_2 \cos(\theta^{\prime} - \theta)\Big) \frac{ \mathbf{x^{\prime}} - \mathbf{x}}{|\mathbf{x^{\prime}}-\mathbf{x}|^2} \Bigg] \rho(\mathbf{x}^{\prime},\theta^{\prime},t) d \mathbf{x^{\prime}} d \theta^{\prime} \\
%v_{\theta} &= K \int \frac{\sin(\theta^{\prime}-\theta)}{|\mathbf{x^{\prime}} - \mathbf{x}|^2}  \rho(\mathbf{x^{\prime}},\theta^{\prime},t) d \mathbf{x^{\prime}} d \theta^{\prime}.
%\end{align}

%%%%%%%%%%%%%%%%%%%%%%%%%%%%%%%%%%%%%%%%%%%%%%%%%%%%%%%%%%%%

\noindent \textbf{Zero divergence condition}. We first investigate Eq.~\eqref{div_vel}. In polar coordinates the continuum expressions for the velocity $\underline{v}$ are

\begin{align}
v_r = & \int \Big(  s\cos (\phi ^{\prime }-\phi )-r  \Big) \Bigg( 1+J_{1}\cos (\theta ^{\prime }-\theta ) \nonumber \\
&- \frac{1 - J_2 \cos(\theta^{\prime} - \theta)}{ s^2 - 2 r s \cos(\theta^{\prime} - \theta) + r^2}    \Bigg)  \;s\rho
(s,\phi ^{\prime },\theta ^{\prime })dsd\phi ^{\prime }d\theta ^{\prime }  \\
v_{\phi} = & \int  s \sin (\phi ^{\prime }-\phi )  \Bigg( 1+J_{1}\cos (\theta ^{\prime }-\theta ) \nonumber \\
&- \frac{1 - J_2 \cos(\theta^{\prime} - \theta)}{ s^2 - 2 r s \cos(\theta^{\prime} - \theta) + r^2}    \Bigg)  \;s\rho
(s,\phi ^{\prime },\theta ^{\prime })dsd\phi ^{\prime }d\theta ^{\prime }  \\
 v_{\theta } =&K\int \frac{\sin (\theta ^{\prime }-\theta )}{
s^{2}-2rs\cos (\phi ^{\prime }-\phi )+r^{2}}\;s\rho (s,\phi ^{\prime
},\theta ^{\prime })dsd\phi ^{\prime }d\theta ^{\prime }.
\end{align}
\noindent
where $v_{\phi} = r \dot{\theta}$. Substituting the ansatz \eqref{density_ansatz} for the density $%
\rho$ into the velocity fields above leads to $v_{\phi} = v_{\theta} = 0$.
The radial component becomes
\begin{align}
v_r =& \frac{1}{2 \pi} \int_{R_1}^{R_2} \int_{-\pi}^{\pi} \Big( s \cos \beta
- r \Big) g(r) s ds d \beta \nonumber \\
&- \frac{1}{2 \pi} \int_{R_1}^{R_2} \int_{-\pi}^{\pi} \frac{ s \cos(\beta) - r}{ s^2 -2 r s \cos \beta + r^2}
g(s) s ds d \beta \nonumber \\
& + \frac{J_1}{2 \pi} \int_{R_1}^{R_2} \int_{-\pi}^{\pi} \Big( s \cos^2 \beta -
r \cos \beta \Big) g(s) s ds d \beta  \nonumber
\\ & + \frac{J_2}{2 \pi} \int_{R_1}^{R_2}
\int_{-\pi}^{\pi} \frac{ s \cos^2 \beta - r \cos \beta}{ s^2 -2 r s
\cos\beta + r^2} g(s) s ds d \beta
\end{align}

\noindent where $\beta = \phi^{\prime }- \phi$. Evaluating the first and
third integrals is elementary, while the second and fourth can be computed
using Poisson's formula,
\begin{equation}
\frac{1}{2 \pi} \int_{-\pi}^{\pi} \frac{\cos m \theta}{s^2 - 2 r
\cos \theta + r^2} d \theta = 
\begin{cases}
(\frac{r}{s})^m \frac{1}{s^2 - r^2} & \text{if } r < s \\ 
(\frac{s}{r})^m \frac{1}{r^2 - s^2} & \text{if } r > s%
\end{cases}%
\end{equation}

\noindent The result is
\begin{align}
v_r &= -r \int_{R_1}^{R_2} g(s) s ds + \frac{1}{r} \int_{0}^{r} s g(s) ds
  + \frac{J_1}{2} \int_{R_1}^{R_2} s^2 g(s) ds  \nonumber \\ &+ \frac{J_2}{2} \int_r^{\infty}
g(s) ds - \frac{J_2}{2 r^2} \int_0^r s^2 g(s) ds.  \label{v_r}
\end{align}
\noindent
In polar coordinates the divergence is
\begin{equation}
\nabla . v = \frac{1}{r} \frac{\partial }{ \partial r}(r v_r) + \frac{1}{r} 
\frac{\partial }{ \partial \phi}( v_{\phi}) + \frac{\partial }{ \partial
\theta}( v_{\theta}).
\end{equation}

\noindent Since $v_{\phi} = v_{\theta} = 0$ this reduces to
\begin{equation}
\nabla . v = \frac{1}{r} \frac{\partial }{ \partial r}(r v_r).
\end{equation}

\noindent Substituting $v_r$ as per \eqref{v_r} into the above expression and applying the derivative operator gives
\begin{align}
\nabla . v &= \frac{1}{r} \Bigg( -2 r \int_{R_1}^{R_2} g(s) s ds + r g(r)(1 -
J_2) \nonumber \\ 
&+ \frac{J_1}{2} \int_0^{\infty} s^2 g(s) ds + \frac{J_2}{2}
\int_r^{\infty} g(s) s^2 ds \nonumber \\
&- \frac{J_2}{2 r^2} \int_0^r s^2 g(s) ds \Bigg).
\end{align}

\noindent Setting this to zero, as required by \eqref{div_vel}, and rearranging, leads to the following
integral equation for $g(r)$
\begin{align}
g(r)&=\frac{1}{1- J_{2}} \Bigg( 2-\frac{J_{1}}{2r}%
\int_{R_{1}}^{R_{2}}s^{2}g(s)ds-\frac{J_{2}}{r}\int_{r}^{R_{2}}g(s)ds \nonumber \\
&-\frac{J_{2}}{r^{3}}\int_{R_{1}}^{r}s^{2}g(s)ds \Bigg)  \label{integral_eqn}.
\end{align}

%%%%%%%%%%%%%%%%%%%%%%%%%%%%%%%%%%%%%%%%%%%%%%%%%%%%%%%%%%%%
\noindent \textbf{Solve integral equation}. We solve the above integral equation for $g(r)$ by reducing it to an ODE. Multiplying both sides by $r^3$ and taking a derivative with respect to $r$ gives
\begin{align}
3 r^2 g(r) + r^3 g^{\prime }(r) &= \frac{1}{1 - J_2} \Bigg[ 6 r^2 - J_2 r
\int_r^{\infty} g(s) ds \nonumber \\
& \hspace{1.5 cm}+ J_1 r \int_{R_1}^{R_2} s^2 g(s) ds \Bigg]
\end{align}

\noindent We next divide by $r$ to give
\begin{align}
3 r g(r) + r^2 g^{\prime }(r) &= \frac{1}{1 - J_2} \Bigg[ 6 r- J_2 
\int_r^{\infty} g(s) ds \nonumber \\
& \hspace{1.5 cm}+ J_1 \int_{R_1}^{R_2} s^2 g(s) ds \Bigg]
\end{align}
\noindent since this expression is easier to differentiate, as there then are only constants in front of the integrals. Taking the derivative then leads  the following simple, second order ODE for $g(r)$
\begin{equation}
r^2 g^{\prime \prime }(r)+5 r g^{\prime }(r)+\left( 3 - \frac{J_2}{1 - J_2}
\right) g(r)-\frac{6}{1- J_2} = 0.  \label{g_r_ode}
\end{equation}

\noindent The solution to this equation is
\begin{equation}
g(r) = C_1 r^{-\frac{1}{\sqrt{1-J_2}}-2}+C_2 r^{\frac{1}{\sqrt{1-J_2}}-2}+\frac{6}{3-4 J_2}.  \label{g_soln}
\end{equation}
We find the constants of integration $C_1, C_2$ by
substituting this back into the integral equation \eqref{integral_eqn}, which gives
\begin{equation}
\frac{A}{r} + \frac{B}{r^3} = 0
\end{equation}
\noindent where $A,B$ are complex functions of $C_{1},C_{2},R_{1},R_{2},J_{1},J_{2}$
that must be identically $0$. Enforcing this constraint leads to the
following complicated expressions for $C_{1},C_{2}$. 

\begin{widetext}
{\small 
\begin{align}
C_{1}& =-\frac{2R_{1}^{\frac{1}{\sqrt{1-J_{2}}}}R_{2}^{\frac{1}{\sqrt{1-J_{2}%
}}}\left( J_{1}\left( \sqrt{1-J_{2}}-1\right) R_{2}^{2}\left(
R_{2}^{2}R_{1}^{\frac{1}{\sqrt{1-J_{2}}}}-R_{1}^{2}R_{2}^{\frac{1}{\sqrt{%
1-J_{2}}}}\right) +J_{2}\left( 3\left( \sqrt{1-J_{2}}-1\right)
R_{2}^{2}R_{1}^{\frac{1}{\sqrt{1-J_{2}}}}+\left( \sqrt{1-J_{2}}+1\right)
R_{1}^{2}R_{2}^{\frac{1}{\sqrt{1-J_{2}}}}\right) \right) }{\sqrt{1-J_{2}}%
\left( 4J_{2}-3\right) \left( \left( -J_{1}R_{2}^{2}+J_{2}+2\sqrt{1-J_{2}}%
-2\right) R_{1}^{\frac{2}{\sqrt{1-J_{2}}}}+R_{2}^{\frac{2}{\sqrt{1-J_{2}}}%
}\left( J_{1}R_{2}^{2}-J_{2}+2\sqrt{1-J_{2}}+2\right) \right) }  \label{C1}
\\
C_{2}& =-\frac{2J_{2}\left( \left( -J_{1}R_{2}^{2}+J_{2}+2\sqrt{1-J_{2}}%
-2\right) R_{1}^{\frac{1}{\sqrt{1-J_{2}}}+2}+R_{2}^{\frac{1}{\sqrt{1-J_{2}}}%
+2}\left( J_{1}R_{2}^{2}+3J_{2}\right) \right) }{\left( \sqrt{1-J_{2}}%
-1\right) \sqrt{1-J_{2}}\left( 4J_{2}-3\right) \left( \left(
-J_{1}R_{2}^{2}+J_{2}+2\sqrt{1-J_{2}}-2\right) R_{1}^{\frac{2}{\sqrt{1-J_{2}}%
}}+R_{2}^{\frac{2}{\sqrt{1-J_{2}}}}\left( J_{1}R_{2}^{2}-J_{2}+2\sqrt{1-J_{2}%
}+2\right) \right) }.  \label{C2}
\end{align}%
}
\end{widetext}

\noindent 
Looking at the third term of the expression for $g(r)$ given by Eq.~\eqref{g_soln}, we see the value
 $J_2 = 3/4$ is problematic. Why is this value distinguished? The reason is that the third term
 in the ODE \eqref{g_r_ode} for $g(r)$ becomes zero at this value of $J_2$. In this case, the ODE 
 has solution 
\begin{equation}
g(r) = -\frac{C_1}{4 r^4}+C_2+6 \log{r}, \hspace{0.5 cm} J_2 = 3/4 \label{g_soln1}
\end{equation}

\noindent
where the constants $C_1, C_2$ are now 
\begin{widetext}
\begin{align}
C_1 &= \frac{8 R_1^4 R_2^4 \Big( \left(4 J_1 R_2^2+9\right) \log{R_1}-\left(4 J_1 R_2^2+9\right) \log {R_2}+6\Big)}{-4 J_1 R_2^6+4 J_1 R_1^4 R_2^2-9
   R_2^4+R_1^4} \\
C_2 &= \frac{2 \Big(-4 J_1 R_2^6+4 J_1 R_1^4 R_2^2+3 R_2^4 \left(4 J_1 R_2^2+9\right) \log{R_2}-3 R_1^4 \left(4 J_1 R_2^2+1\right) \log{R_1}-27
   R_2^4+R_1^4\Big)}{-4 J_1 R_2^6+4 J_1 R_1^4 R_2^2-9 R_2^4+R_1^4}
\end{align}
\end{widetext}

\begin{figure*}[t]
\includegraphics[width= 1.8 \columnwidth]{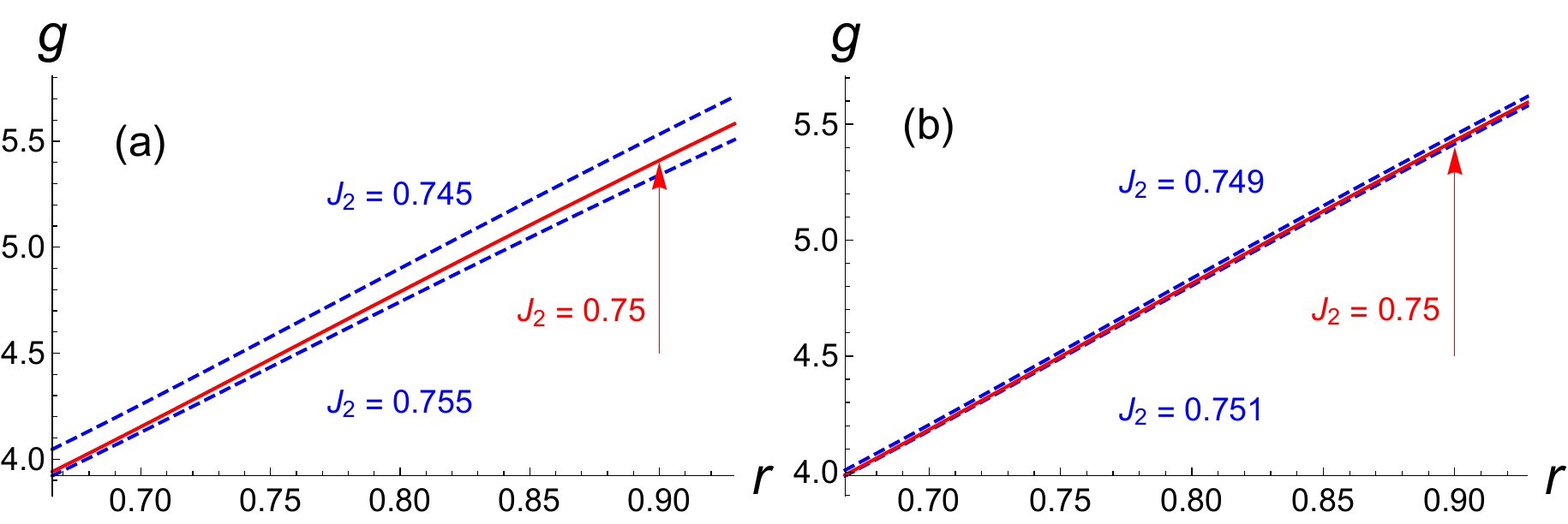}
\caption{Radial density $g(r)$ for $J_1 = 0.5$ for values of $J_2$ in a neighborhood
of $0.75$. Blue dashed lines show are for $J_2 \neq 3/4$ calculated using expression \eqref{g_soln}.
The red solid line is for $J_2 = 3/4$ using expression \eqref{g_soln1}. The density $g(r)$ varies
smoothly as $J_2$ passes through $0.75$. Panel (a) shows values $J_2 = 0.745, 0.755$, which hug
the curve at $J_2 = 0.75$. In panel (b) we use a tighter neighborhood with extremal values $0.749, 0.751$,
which produces a tighter `hugging'. These results indicate that there is no change in the behavior of $g(r)$ at the value $J_2  = 0.75$. }
\label{g_smooth}
\end{figure*}

The difference between the expressions \eqref{g_soln} and \eqref{g_soln1} for $g(r)$ are superficial. By this we mean
there is no change in the physical behavior of the swarmalator system as $J_2$ passes through $3/4$. We demonstrate
this two ways. The first way is by observing that $R_1, R_2$ vary smoothly with respect to $J_2$ as drawn in Figure~\ref{radii_of_annular_distribution}; no change in behavior occurs at $J_2 = 3/4$. The second way is by plotting 
$g(r)$ at the values for values of $J_2$ is the neighborhood of $3/4$ in Figure~\ref{g_smooth}. As can be seen $g(r)$ varies smooth as $J_2$ is varied through $3/4$. Hence the value of $J_2 = 3/4$ is a mathematical quirk,
and has no physical significance.

%% Figure 6

%%%%%%%%%%%%%%%%%%%%%%%%%%%%%%%%%%%%%%%%%%%%%%%%%%%%%%%%%%

\subsection{Inner and outer radii}
So far we have solved for $g(r)$ using the zero divergence condition \eqref{div_vel}. The zero velocity condition $\eqref{vel}$ must also be satisfied. We here check the condition $v_r  = 0$, and show that along with mass conservation $\int \rho(x,\theta) dx d\theta = 1$, it also lets us determine the inner and outer radii $R_1, R_2$. \\

\noindent
\textbf{Zero velocity condition}
Substituting the expression \eqref{g_soln} for $g(r)$ into Eq.~\eqref{v_r} for $v_r$ leads to 
\begin{equation}
v_r = \frac{h_1(R_1, R_2, J_1, J_2)}{r} 
\end{equation}

\noindent
where $h_1$ is given by

\begin{widetext}
\begin{align}
h_1 &= \Bigg[2 J_2^2 \left(2 \sqrt{J_2+1}+3 R_2^2-6\right)+J_2 \left(R_2^2 \left(J_1 \left(4 \sqrt{J_2+1}-2 R_2^2-4\right)-15 \sqrt{J_2+1}+21\right) +19 \sqrt{J_2+1}-25\right) \nonumber \\
&+\left(\sqrt{J_2+1}-1\right) \left(J_1 R_2^4+3 \left(J_1-4\right) R_2^2+12\right)\Bigg] R_1^{\frac{2}{\sqrt{J_2+1}}} 
   +4 \left(J_2-\sqrt{J_2+1}+1\right) R_2^{\frac{1}{\sqrt{J_2+1}}} \left(J_1 R_2^2+J_2\right) R_1^{\frac{1}{\sqrt{J_2+1}}+2} \nonumber \\
   &+\Bigg[ J_2 \left(-2 J_1 R_2^2+7
   \sqrt{J_2+1}-13\right)+3 \left(\sqrt{J_2+1}-1\right) \left(J_1 R_2^2+4\right)-2 J_2^2\Bigg] R_1^{\frac{2}{\sqrt{J_2+1}}+2} \nonumber \\
   &+4 \left(-J_2+\sqrt{J_2+1}-1\right) R_2^{\frac{1}{\sqrt{J_2+1}}+2} \left(3 J_2-J_1 R_2^2\right) R_1^{\frac{1}{\sqrt{J_2+1}}} \nonumber \\
   &-R_1^2 R_2^{\frac{2}{\sqrt{J_2+1}}} \Bigg[J_2 \left(2 J_1 R_2^2+3
   \sqrt{J_2+1}+3\right)-J_1 \left(\sqrt{J_2+1}-1\right) R_2^2+2 J_2^2\Bigg] \nonumber \\
   &-R_2^{\frac{2}{\sqrt{J_2+1}}} \Bigg[J_2^2 \left(4 \sqrt{J_2+1}-6 R_2^2+4\right)+J_2
   \left(R_2^2 \left(2 J_1 \left(2 \sqrt{J_2+1}+R_2^2-2\right) -3 \left(\sqrt{J_2+1}+1\right)\right) 
   +3 \left(\sqrt{J_2+1}+1\right)\right) \nonumber  \\ 
   &-3 J_1
   \left(\sqrt{J_2+1}-1\right) R_2^2 \left(R_2^2-1\right) \Bigg]. \label{h1}
\end{align}
\end{widetext}

\noindent
We require $v_r  = 0$ for all $r$, which implies $h_1(R_1,R_2,J_1,J_2) = 0$. \\  

\noindent
\textbf{Mass conservation}. The density ansatz \eqref{density_ansatz} must also be normalized: $\int \rho(x,\theta) dx d \theta = 1$. This leads to a second equation $h_2(R_1, R_2, J_1, J_2) = 0$ where
\begin{widetext}
\begin{align}
h_2 &= - \Bigg[ J_2 \left(2 J_1 R_2^2+3 \sqrt{J_2+1}+3\right)-J_1 \left(\sqrt{J_2+1}-1\right) R_2^2+2 J_2^2 \Bigg] R_1^2 \; R_2^{\frac{2}{\sqrt{J_2+1}}} \nonumber \\
 &+4 \left(-J_2+\sqrt{J_2+1}-1\right) R_1^{\frac{1}{\sqrt{J_2+1}}} \left(3 J_2-J_1 R_2^2\right) R_2^{\frac{1}{\sqrt{J_2+1}}+2} \nonumber \\
 &+ \Bigg[J_2
   \left(-2 J_1 R_2^2+7 \sqrt{J_2+1}-13\right)+3 \left(\sqrt{J_2+1}-1\right) \left(J_1 R_2^2+4\right)-2 J_2^2\Bigg] R_1^{\frac{2}{\sqrt{J_2+1}}+2}.  \label{h2}
\end{align}
\end{widetext}
\noindent
\\ \\
Thus we have derived the following set of simultaneous equations whose roots
determine $R_1, R_2$ in terms of the parameter $J_1$ and $J_2$.
\begin{align}
h_1(R_1, R_2, J_1, J_2) = 0   \\
h_2(R_1, R_2, J_1, J_2) = 0.
\end{align}

%%%%%%%%%%%%%%%%%%%%%%%%%%%%%%%%%%%%%%%%%%%%%%%%%%%%%%%%%%

%%%%%%%%%%%%%%%%%%%%%%%%%%%%%%%%%%%%%%%%%%%%%%%%%%%%%%%%%%

\section{Genericity}

%% Figure 7
\begin{figure*}[t]
\includegraphics[width= 1.5 \columnwidth]{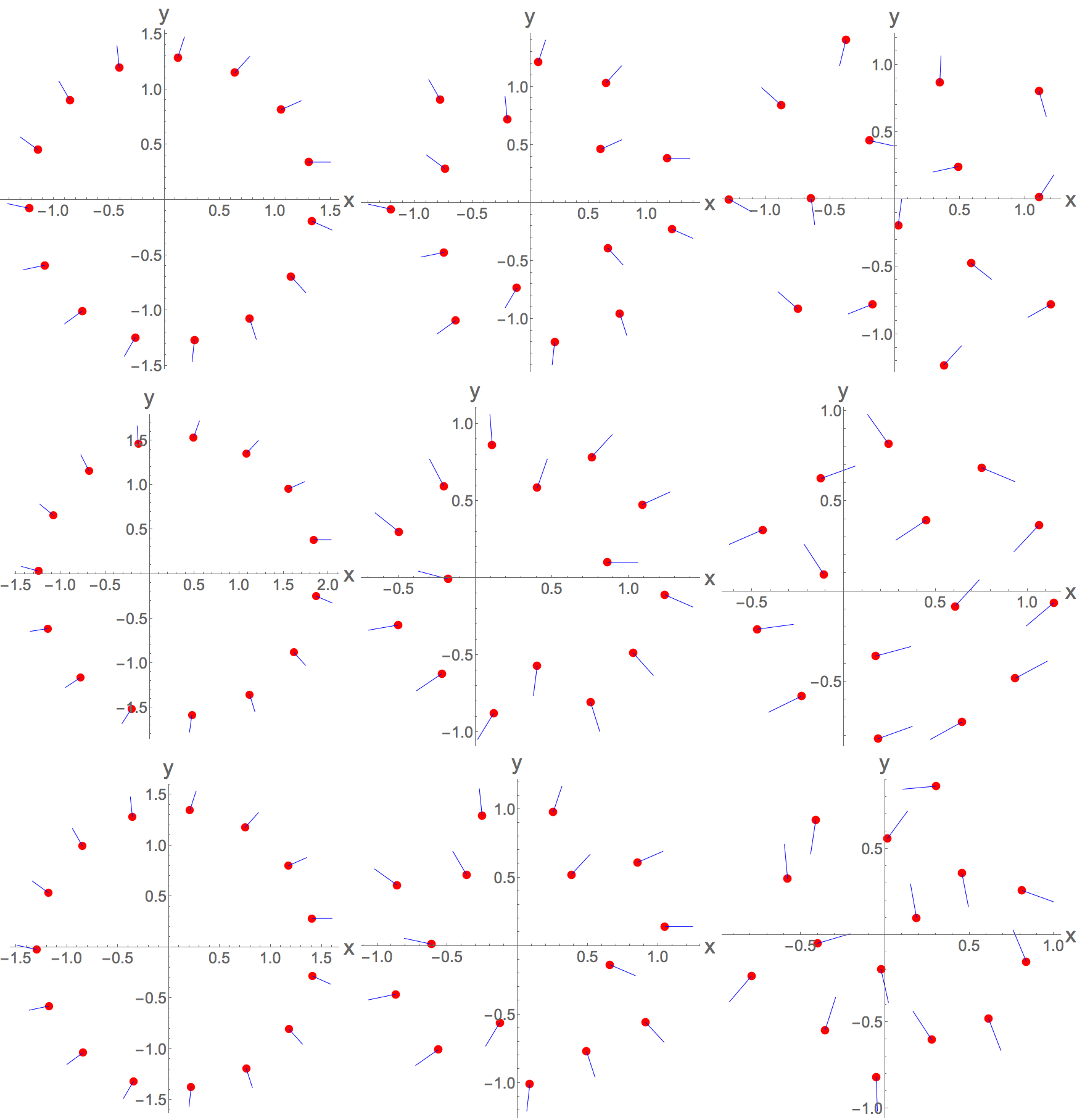}
\caption{States found with difference choices of the functions \eqref{gener1}, \eqref{gener2} and \eqref{gener3}. Simulations for all plots were for $N = 15$ swarmalators, and the Euler method with a stepsize of $dt = 0.01$ and $N_t = 5 \times 10^5$ number of timesteps was used. The top row is for choice \eqref{gener1}, the second for choice \eqref{gener2}, and the third for choice \eqref{gener3}. The ring state, corresponding to subfigure B in the stability diagram in Figure~\ref{fig:stab}, is shown in the first column. Reading from top to bottom, the parameter values were $(J_1, J_2, K) = (2.7, 0, -0.001), (1,0, 0, - 0.01), (1.5, 0, -0.001)$. The fattened ring state, corresponding to subfigure E in Figure~\ref{fig:stab}, is shown in the second column. Parameter values were $(J_1, J_2, K) = (1.5, 0, -0.001), (0.2,0, 0, - 0.01), (0.8, 0, -0.001)$. The column shows the non-stationary state depicted in subfigure G in Figure~\ref{fig:stab}. Parameter values were $(J_1, J_2, K) = (1.5, 0, -5), (0.2,0, 0, -2), (0.8, 0, 5)$. Note, in this last column, the swarmalators move around erratically in both space and phase.}
\label{genericity}
\end{figure*}

\end{document}